\documentclass[useAMS,usenatbib,usegraphicx]{mn2e}
\usepackage[utf8]{inputenc}

\usepackage[usenames]{color}
\usepackage{ulem}

\usepackage{graphicx}	

\newcommand{\kms}   {km~s$^{-1}$}

\newcommand{\cmt}   {cm$^{-3}$}

\newcommand{\nh}    {NH$_3$}

  \def\alamenos#1{$^{-#1}$}

  \def\diezala#1{10$^{#1}$}
  \def\Msun{{$M_\odot$}}

  \def\larsonratio{$\sigma_{v, {\rm 3D}}/R^{1/2}$}
  
  \def\larsonratiodisp{\frac{\sigma_{\rm v, 3D}} {R^{1/2}}}
  \def\sigmatd{{$\sigma_{{\rm v,  3D}}$}}
  \def\sigmatdmath{{\sigma_{v,{\rm 3D}}}}

  \newcommand{\dvr}{\sigmatd$-R$}

  \newcommand{\dvrSigma}{$\LL$--$\Sigma$}

  \def\ltsima{$\; \buildrel < \over \sim \;$}    
  \def\lesssim{\lower.5ex\hbox{\ltsima}}           
  \def\gtsima{$\; \buildrel > \over \sim \;$}    
  \def\gtrsim{\lower.5ex\hbox{\gtsima}}           

  \newcommand{\beq}{\begin{equation}}
  \newcommand{\bfig}{\begin{figure}}
  
  \newcommand{\eeq}{\end{equation}}
  \newcommand{\efig}{\end{figure}}
  \newcommand{\Eg}{E_{\rm g,sph}}
  \newcommand{\Ek}{E_{\rm k}}
  \newcommand{\Etot}{E_{\rm tot}}
  \newcommand{\Lff}{{\cal L}_{\rm ff}}
  \newcommand{\LL}{{\cal L}}
  \newcommand{\Lin}{{\cal L}_{\rm in}}
  \newcommand{\Ltot}{{\cal L}_{\rm tot}}
  \newcommand{\sigtd}{\sigma_{v,{\rm 3D}}}
  \newcommand{\sigtot}{\sigma_{\rm tot}}
  
  \newcommand{\tff}{\tau_{\rm ff}}
  \newcommand{\vff}{\sigma_{\rm ff}}
  \newcommand{\vin}{\sigma_{\rm in}}
  \newcommand{\vturb}{\sigma_{\rm turb}}

  \def\ltsima{$\; \buildrel < \over \sim \;$}    
  \def\lesssim{\lower.5ex\hbox{\ltsima}}           
  \def\gtsima{$\; \buildrel > \over \sim \;$}    
  \def\gtrsim{\lower.5ex\hbox{\gtsima}}           

\def\apss{Ap\& SS}
\def\apj{ApJ}
\def\apjl{ApJL}
\def\aap{A\& A}

\def\araa{ARA\&A}
\def\mnras{MNRAS}

\def\apjs{ApJS}

\def\mnras{{MNRAS}}

\title[Short title, max. 45 characters]{MNRAS \LaTeXe\ template -- title goes here}

\title[Collapsing cores in out-of-virial disguise]{Gravity or turbulence?
  IV. Collapsing cores in out-of-virial disguise}
\author[Ballesteros-Paredes et al.] {Javier
  Ballesteros-Paredes$^{1,2}$\thanks{E-mail:j.ballesteros@crya.unam.mx},
  Enrique V\'azquez-Semadeni$^{1}$, Aina
  Palau$^1$, Ralf S. Klessen${^{2,3}}$  %
  \\ $^{1}$ Instituto de Radioastronom\'ia y Astrof\'isica,
  Universidad Nacional Aut\'onoma de M\'exico, P.O. Box 3-72, 58090
  Morelia, Michoac\'an, M\'exico\\ $^{2}$ Universit\"{a}t Heidelberg,
  Zentrum f\"{u}r Astronomie, Institut f\"{u}r Theoretische
  Astrophysik, Albert-Ueberle-Straße 2, 69120 Heidelberg, Germany
  \\ $^3$Universit\"{a}t Heidelberg, Interdisziplin\"{a}res Zentrum
  f\"{a}r Wissenschaftliches Rechnen, Im Neuenheimer Feld 205, 69120
  Heidelberg, Germany
}

\date{Accepted XXX. Received YYY; in original form ZZZ}

\pubyear{2015}

\begin{document}
\label{firstpage}
\maketitle

\begin{abstract}
We study the dynamical state of cores by using a simple analytical
model, a sample of observational massive cores, and numerical
simulations of collapsing massive cores.  From the analytical model,
we find that, if cores are formed from turbulent compressions, they
evolve from small to large column densities, increasing their velocity
dispersion as they collapse. This results in a time evolution path in
the Larson velocity dispersion-size diagram from large sizes and small
velocity dispersions to small sizes and large velocity dispersions,
while they tend to equipartition between gravity and kinetic energy.

From the observational sample, we find that: (a) cores with
substantially different column densities in the sample do not follow a
Larson-like linewidth-size relation. Instead, cores with higher column
densities tend to be located in the upper-left corner of the Larson
velocity dispersion \sigmatd-size $R$ diagram, a result predicted
previously \citep{BP11a}.  (b) The data exhibit cores with overvirial
values.

Finally, in the simulations of collapsing cores we reproduce the
behavior predicted by the analytical model and depicted in the
observational sample: cores evolve towards larger velocity dispersions
and smaller sizes as they collapse and increase their column
density. More importantly, however, is that collapsing cores appear to
approach overvirial states within a free-fall time.  We find that the
cause of this apparent excess of kinetic energy is an underestimation
of the actual gravitational energy, due to the assumption that the
gravitational energy is given by the energy of an isolated sphere of
constant column density.  We find that this apparent excess disappears
when the gravitational energy is correctly calculated from the actual
spatial mass distribution, where inhomogeneities, as well as the
potential due to the mass outside of the core, also contribute to the
gravitational energy.  We conclude that the observed energy budget of
cores in recent surveys is consistent with their non-thermal motions
being driven by their self-gravity and in the process of dynamical
collapse.

\end{abstract}





\date{Accepted date. Received date; in original form date}


\maketitle


\section{Introduction}\label{sec:intro}

Since the first detections of molecular gas, it was recognized that
their line profiles exhibit supersonic widths (Wilson et al.  1970).
Early models interpreting such profiles suggested that they were the
signature of clouds in a state of large-scale collapse, since
turbulence should decay over a dynamical crossing time
\citep{GoldreichKwan74, Liszt+74}. This proposal was rapidly dismissed
by \citet{ZuckermanPalmer74}, who argued that the star formation rates
would be too large if clouds were in a state of {free-fall}.  These
authors proposed, instead, that the supersonic linewidths were
produced by small-scale turbulence, which furthermore, {might} provide
support to molecular clouds (MCs) against their own collapse
\citep[see, e.g., ][and references therein]{VS+00, MK04, BP07,
  McKeeOstriker07, KlessenGlover16}.

Indeed, there is a number of apparently good reasons to believe that
turbulence may play an important role in the structure and dynamics of
MCs. First, the Reynolds numbers in the interestellar medium are very
large, indicating that the velocity field should be strongly turbulent
\citep{ElmegreenScalo04}.  Second, the velocity field in the ISM
exhibits a multi-scale nature that extends over several orders of
magnitud in size \citep[e.g., ][]{Larson81}.  Third, clouds exhibit a
fractal appareance \citep{Falgarone+91}, characteristic of turbulent
fluids.  Fourth, stellar jets and winds, gravitational chaotic motions
of orbiting gas in the galaxy, SN explosions, spiral arm shocks, etc.,
may be powering the gas of kinetic energy at multiple scales
simultaneously \citep[e.g.,] [] {NF96}.  In addition, the large widths
of the observed spectral lines indicate that such turbulence should be
supersonic, reinforcing the idea that turbulence is intrinsically
related to MC structure by shaping, morphing and fragmenting MCs
\citep{BP99a, Hopkins12}. {In the prevailing interpretation}, MCs are
thought to be near virial equilibrium, with perhaps a moderate excess
of kinetic energy, such that some external pressure is needed to
confine them for several dynamical times \citep{BertoldiMcKee92,
  Field+11, Miville-Deschenes+16, Leroy+15}.  MC turbulence plays a
crucial role providing global support against collapse, while
promoting local collapse where motions are converging \citep[e.g.,] []
{BP99a, MK04, BP06, BP07, HF12, Padoan+14}.

Although the turbulent nature of the velocity field in MCs is hardly
questionable, several problems remain with the standard interpretation
that it can provide support against the global collapse of the clouds:
First, turbulence consists of a hierarchy of coherent motions over a
wide range of scales for which the largest-amplitude velocity
fluctuations occur at the largest scales.  This implies that the
dominant MC turbulent motions are far from being the small-scale
turbulent motions that can provide the internal turbulent pressure to
support the cloud against gravity \citep[see, e.g.,] [] {BP06}, as
confirmed numerically by \citet{Brunt+09}.  Thus, the dominant motions
induced by turbulence at the scale of clouds will be cloud-scale
motions such as compression, expansion, shearing or rotation, but not
thermal-like, small-scale motions that can provide an internal
pressure to support it against gravity \citep[see, e.g.,] [] {BP99a,
  BP06, VS15}.

Second, since turbulence is a dissipative phenomenon, it requires
constant driving to be maintained.  The standard belief is that the
turbulent energy is injected by mainly two mechanisms: one, through
instabilities occurring during the assembly stage of the clouds
\citep[e.g.,] [] {Vishniac94, WF00, KI02, AH05, Heitsch+05, VS+06,
  KH10}.  The other, by stellar feedback via outflows, winds, ionizing
radiation or supernova explosions \citep[e.g.,] [] {MK04, Wang+10,
  VS+10, Gatto+15, Padoan+16}. Regarding the first point, numerical
simulations of the formation of MCs from converging flows generally
suggest that the turbulence level produced by various instabilities is
significantly lower ($\la 20\%$) than the level typically observed in
MCs \citep[e.g.,] [] {KI02, Heitsch+05, AH05, VS+07, VS+10}, even with
the inclusion of external SN explosions that trigger the compressions
which form the clouds \citep{Ibanez+16}, although this is still matter
of debate \citep[e.g., ][]{Padoan+16}.  With respect to the stellar
feedback, it is not {clear why} the energy it injects would cause
apparent near virialization of {\it all} molecular structures. In
fact, numerical simulations suggest quite the opposite: clumps and
moderate-mass MCs appear to be readily disrupted by photoionizing
radiation or supernova explosions from massive stars, while high-mass
MCs are difficult to prevent from collapsing \citep[e.g.,] []
{Dale+12, Dale+13a, Dale+13b, Colin+13}.  Thus, neither of the two
mechanisms is likely to produce the observed magnitude of the
nonthermal motions in MCs.

On the other hand, in the last decade, numerical simulations of the
formation and evolution of molecular clouds have renewed the idea that
molecular clouds may be in a state of collapse.  These models suggest
that their formation from convergent motions in the warm neutral
medium (WNM) in the Galaxy \citep{BP99a, BP99b} involves a phase
transition to the cold neutral medium \citep[CNM;] [] {HP99, KI00,
  VS+06}, and thus the clouds may be formed with masses much larger
than their Jeans mass \citep{HBB01, VS+07, HH08, GVS14}, implying that
they may very well be in a state of large-scale hierarchical and
chaotic collapse \citep{VS+09, BP11a}.

Note, however, that, contrary to the suggestion by
\citet{ZuckermanPalmer74}, such a state of global collapse does not
necessarily imply that the star formation rate will be too large: The
initial fragmentation occuring while the cloud is being assembled has
the consequence that only very small amounts of mass are deposited in
the high density regions that are responsible for the instantaneous
rate of star formation in the clouds, as evidenced by the steep
negative slope of the column density probability distribution
functions of MCs \citep[][]{Kainulainen+09, BP11b, Kritsuk+11}.
Having large nonlinear amplitudes, such dense regions complete their
collapse much earlier than the whole cloud, causing a spread in the
ages of the stellar products \citep[e.g., ] [] {Zamora+12,
  Hartmann+12, VS+17}.  This will eventually produce massive stars
that will blow strong winds and rapidly ionize the gas, effectively
dispersing the nearby dense gas \citep[e.g., ] [] {VS+10, Zamora+12,
  Dale+12, Dale+13a, Dale+13b, Colin+13, ZV14, VS+17}, shutting off
the local star formation episodes by the time only $\sim 10\%$ of the
gas has been converted to stars, thus keeping the global star
formation rate and efficiency low.

 In this scenario, the supersonic nonthermal motions observed in MCs,
 albeit still turbulent in the sense of having a chaotic component,
 are dominated by an infall component that occurs at multiple scales,
 consituting a regime of collapses within collapses \citep[e.g., ] []
 {VS+09, VS+17}. This implies that the nonthermal motions are
 dominated by a convergent component that {results from} the
 gravitational contraction, rather than being a fully random velocity
 field that can provide a pressure gradient capable of opposing the
 collapse \citep{VS+08, Gonzalez-Semaniego+14}. Nevertheless, the chaotic,
 moderately-supersonic nature of the truly turbulent motions does
 produce a spectrum of density fluctuations that provide the focusing
 centers for the multi-scale collapse \citep{CB05}.

 {Two important pieces of evidence supporting the scenario of global,
   hierarchical collapse are worth noting.} On { the} theoretical
 side, as \citet{VS+07} showed, during the early {stages of the}
 formation of a MC, the kinetic energy, {driven by instabilities in
   the dense layer produced by the collision of diffuse gas strams,} {
   is not coupled to} the gravitational energy. The {two energies}
 become correlated once {the cloud becomes dominated by gravity and
   begins to collapse}.  (see Fig.\ {8 of} \citet{VS+07}).  In other
 words, {the gravitational} collapse is the physical agent that
 {induces apparent} virialization between the energies of the cloud.

On the observational side, clumps with sufficiently different column
densities do not conform to a unique Larson-like velocity
dispersion-size relation.  Instead, it has been shown by different
authors \citep{Heyer+09, BP11a, Leroy+15, Traficante+16} that when the
sample objects span a large enough column density dynamic range, then
they follow a relation of the form

\begin{equation}
  \sigma_{NT} \approx \sqrt{G \Sigma R}.
  \label{eq:larson_general}
\end{equation}
where $\sigma_{\rm NT}$ is the velocity dispersion of the non-thermal
motions, $\Sigma$ the column density and $R$ the size of the core, and
$G$ the universal gravitation constant.  This relation explains in a
self-consistent way the two former \cite{Larson81} relations: if {by
  definition,} surveys tend to select objects of roughly constant
column densities, the Larson density-size relation is trivially
satisfied, and then relation (\ref{eq:larson_general}) with a constant
column density implies that the velocity dispersion scales
approximately with size as $R^{1/2}$.  The reason {why} MCs exhibit
frequently nearly constant column densities {in surveys} \citep[e.g.,
][]{Larson81, Solomon+87, Kauffmann+10a, Kauffmann+10b, Lombardi+10,
  Roman-Duval+10} is because most of the area of the clouds is at low
column densities, since the column density probability distribution
function (N-PDF) decreases abruptly {at high column density} , and
thus the mean column density over the projected area of a cloud
defined at a given column density threshold has to be close to the
threshold value \citep{BM02, BDH12, Beaumont+12}.  But, if the surveys
have a wide column density dynamic range \citep[e.g.,] [] {Heyer+09,
  Leroy+15, Traficante+16}, or if several surveys of objects of
clearly distinct column densities are combined \citep[e.g.,] []
{BP11a}, then relation (\ref{eq:larson_general}) holds.  Furthermore,
as discussed by \citet{BP11a}, this relation is consistent with both
virial equilibrium and free-fall, since in both cases the velocity
dispersion has a gravitational origin, and moreover, the difference in
the velocity magnitude between the two cases is only a factor of
$\sqrt{2}$.  Thus, relation (\ref{eq:larson_general}) constitutes a
generalization of Larson's relations to the case when $\Sigma$ is not
constant {among the cloud sample}.

The question of whether clouds are in a state of global, hierarchical,
chaotic collapse or instead they are globally supported by turbulence
against collapse for several free-fall times is crucial to the
understanding of the actual dynamics of MCs, to what defines their
internal structure, and to how MCs evolve and form stars. In this
contribution we present new observational and numerical evidence that
massive dense cores may be in a state of collapse, even though often
they may appear to have an excess of kinetic energy.  In
\S\ref{sec:TV_Larson_Heyer} we discuss the evolution of a contracting
core in the Larson diagram and its energy budget. In \S\ref{sec:data}
we present the observational and numerical data, while in
\S\ref{sec:results} we show that the locus of our observational sample
of is similar to the locus of our collapsing cores in the simulations
either in the Larson velocity dispersion-size diagram, as well as in
the Larson's ratio-column density diagram.  In \S\ref{sec:correction}
we show that the aparently overvirial collapsing cores are actually
not overvirial when the shole distribution of mass is taken into
account on the gravitational potential.  Finally, in
\S\ref{sec:discussion} and \ref{sec:conclusions} we make a general
discussion and provide our conclusions, respectively.

\section{Scaling relations and the evolution of turbulent cores}
\label{sec:TV_Larson_Heyer}

\subsection{Larson's Relations and Virial Balance}

As discussed in Sec.\ \ref{sec:intro}, there is abundant evidence that
the classical Larson relations consitute a particular case, valid for
samples of objects of roughly the same column density, of the more
general relation presented by \citet{KM86} and \citet{Heyer+09}, who
plotted the ratio \larsonratio\ {\it versus} the column density of
clouds for a wide range of column densities. We will call this
coefficient the {\it Larson ratio} $\LL$; i.e.,

\begin{equation}
  \LL \equiv \larsonratiodisp.
  \label{eq:LL}
\end{equation}
\citet{KM86} noted that high latitude clouds have an excess of kinetic
energy, and {concluded that they} require an external confining medium
to be in equilibrium. More recently, \citet{Heyer+09} analyzed clouds
defined with the area within the half-power isophote of the peak
column density value within the cloud.  {For this sample}, they found
that the Larson ratio correlates with the column density {as}
$\LL\propto \Sigma^{1/2}$, in agreement with
eq.\ (\ref{eq:larson_general}). They interpreted this result as
{implying that the} clouds are in self-gravitational equilibrium.

It is worth {noting} that the sample of \citet{Heyer+09} exhibits, at
face value, {systematic} kinetic energy excesses {with respect to the
  virial value}. These authors argued that {this might be due to the
  fact that the cloud} masses might be {systematically} underestimated
by a factor of $\sim 2$. However, others have suggest{ed instead} that
{the} excess {in} the Larson ratio {must be indicative of the presence
  of} an external pressure {that confines the clouds} \citep{KM86,
  Field+11, Leroy+15}.  We should stress, however, that pressure
confinement is actually quite unlikely in the interior of molecular
clouds. As extensively discussed by \citet{VS+05}, for a density
fluctuation (generically, a ``clump'') to attain a {stable}
equilibrium it is necessary that the object is confined by a tenuous,
warm medium that { provides pressure without adding weight}. This does
not appear feasible in the deep interiors of MCs, where the medium is
roughly isothermal, and there is no diffuse confining phase.
Moreover, some of the highest-column density clumps would require
enormous confining pressures, $\sim 10^{4-9}$ K~\cmt\ \citep{Field+11,
  Leroy+15}, which appear highly unlikely to be attained by the low
density regime inside MCs.

{Instead, a} {much simpler} mechanism for {clumps} to exhibit a
relation like eq. (\ref{eq:larson_general}) is if {they} are
collapsing. {In this case, t}here is no need to replenish turbulence
{nor} to invoke pressure confinement {since, as shown by
  \citet{BP11a}, the velocities produced by gravitational collapse
  necessarily satisfy relation (\ref{eq:larson_general}). }

However, an important question remains: why do most clouds then
exhibit a correlation between this ratio and the column density,
although often with apparent excesses of kinetic energy
\citep{Heyer+09, Leroy+15}, and sometimes with a defficiency of
kinetic energy \citep[e.g.,] [] {Kauffmann+13, Ohashi+16,
  Sanhueza+17}? In the remainder of this contribution we address this
question.

\subsection{{Energy balance evolution of collapsing cores}}
\label{sec:core_evol}

To illustrate the expected evolution of a collapsing core in the
{Larson ratio-column density (\dvrSigma) diagram}, in this section we
consider a spherical, uniform collapsing core that may contain both a
turbulent ($\sigtd$) and a gravitationally-driven (or ``infall'',
$\vin$) components of the velocity dispersion.

It should be noted, however, that the turbulent component we consider
is not necessarily assumed to provide support. Quite the contrary, as
mentioned in Sec.\ \ref{sec:intro}, turbulence is known to have the
largest velocities at the largest scales, and so the dominant
turbulent motions in any structure must be those at the scale of the
structure itself. Generally, these can be compressive, shearing, or
vortical motions \citep{VS+96, VS+08, Federrath+08,
  Gonzalez-Semaniego+14}. Recently, \citet{Camacho+16}, examining
numerical simulations of the formation and collapse of MCs, have found
that clumps that exhibit an excess of the Larson ratio have, in
roughly half of the cases, a negative average velocity
divergence---i.e., a convergence. This implies that these clumps are
being {\it assembled} by external compressive motions that are not
driven by the self-gravity of the cloud, but rather constitute
large-scale turbulent motions in the ambient ISM. The most probable
velocity gradients corresponding to convergence observed by
\citet{Camacho+16} are in the range $-0.5 \la \nabla \cdot {\bf v} \la
0$ \kms\ pc$^{-1}$. It is this type of non-self-gravitating motions
that we have in mind when we consider turbulence in this section. We
now consider how the relative contribution of these motions compare
with those driven by the self gravity of the clouds.

\subsubsection{Core with no turbulence}
\label{sec:non-turb_case}

Let us assume spherically symmetric core that for some reason loses
support and starts to collapse at some time $t_0$, at which time it
has an initial radius $R_0$. Neglecting thermal and magnetic energies,
the total energy of the core can be readily calculated at any time
during contraction as

\beq
\Ek + \Eg = \Etot,
\label{eq:cons_energy}
\eeq
where $\Ek = 1/2\, M \vin^2$, $\Eg =- 3/5\, GM^2/R$, and $\Etot =
-3/5\, GM^2/R_0$ is the total energy, which equals the potential
energy of the core before it starts to collapse. Equation
(\ref{eq:cons_energy}) then becomes
\beq
\vin = \sqrt{\frac{6}{5} GM \left(\frac{1}{R} - \frac{1} {R_0} \right)
}.
\label{eq:vinf}
\eeq
This equation shows that the velocity dispersion associated to the
infall of a collapsing core that initiates its collapse at a finite
 time $t=t_0$ with a radius $R_0$ is always smaller than the free-fall
speed, and asymptotically approaches this speed from below.  

Equation (\ref{eq:vinf}) can be written for the Larson ratio in terms
of the initial ($\Sigma_0$) and instantaneous ($\Sigma$) column
densities of the core as
\begin{eqnarray}
  \Lin \equiv \frac{\vin}{R^{1/2}} &=& \sqrt{\frac{6 \pi} {5} G
    \left(\Sigma - \frac{M} {\pi R R_0} \right)}\nonumber \\
  &=& \sqrt{\frac{6 \pi} {5} G\Sigma \left[1 -
      \left(\frac{\Sigma_0}{\Sigma}\right)^{1/2} \right]},
\label{eq:Lin}
\end{eqnarray}

In contrast, the standard free-fall velocity dispersion $\vff$ is
derived from the condition $\Ek = |\Eg|$ rather than
eq.\ (\ref{eq:cons_energy}), and equals $\sqrt{2}$ times the virial
velocity. The corresponding value of the $\LL$ ratio is
\beq 
\Lff \equiv \frac{\vff}{R^{1/2}} = \sqrt{\frac{6 \pi} {5} G\Sigma}.
\label{eq:Lff}
\eeq

\begin{figure*}
\includegraphics[width=0.45\textwidth]{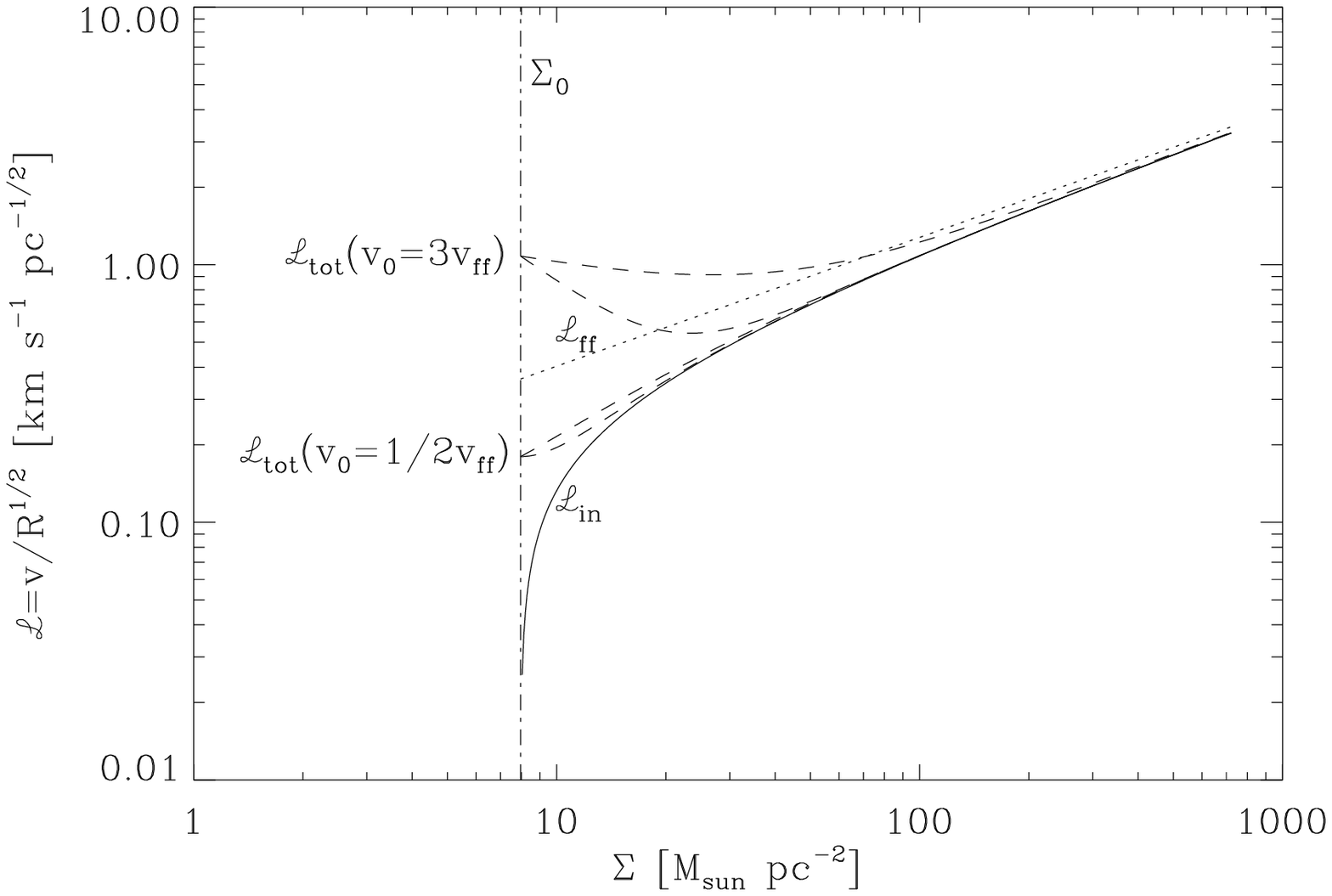}
\includegraphics[width=0.45\textwidth]{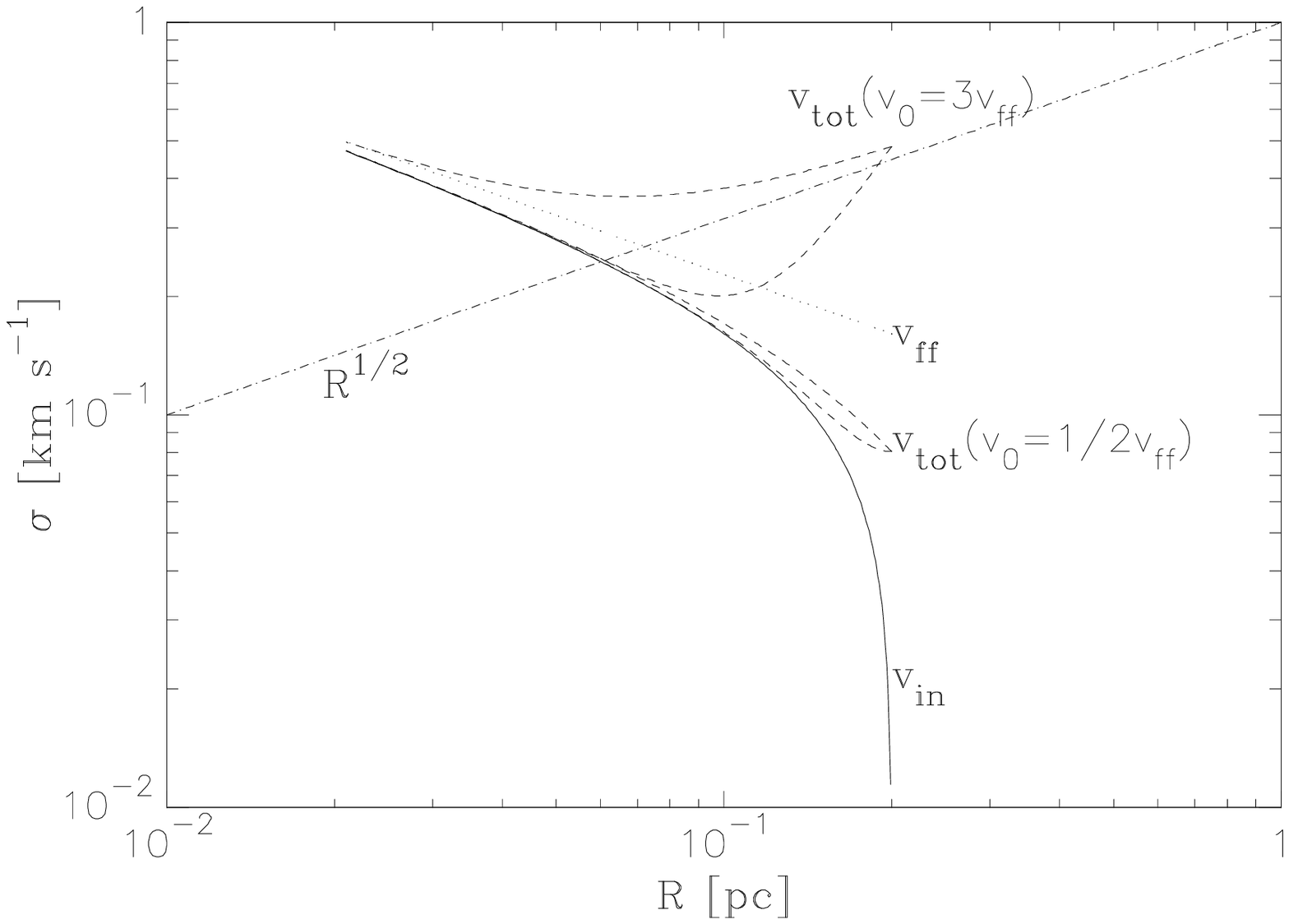}
\caption{{\it Left panel:} The {\it solid line} shows the trajectory in
the $\LL$ {\it vs.} $\Sigma$ plane of a core of fixed mass $M = 1$
\Msun\ that becomes Jeans-unstable, beginning to collapse at a time
$t_0$, at which it has an initial radius $R_0 = 0.2$ pc, implying an
initial column density $\Sigma_0$, shown by the vertical dashed-dotted
line. The {\it dotted line} shows the locus of a core of the same mass
that contracts having always the free-fall speed, as given by eq.\
(\ref{eq:Lff}). The {\it dashed lines} show the evolutionary paths of 4
cores with a combined turbulent+gravitational velocity, as described by
eq.\ (\ref{eq:Ltot}). The upper (resp. lower) pair of dashed lines
correspond to $v_0 = 3 \vff$ (resp.\ $v_0 = 1/2 \vff$).  Of each of
these pairs, the upper curve corresponds to $\eta = 1/2$, appropriate
for a Burgers spectrum, and the lower one to $\eta = 2$, loosely
representing dissipation in dense objects. {\it Right panel} The
evolution of the same set of cores (with the same labeling of the lines)
in the Larson diagram, \sigmatd\ {\it vs.} $R$, represented by the {\it
dashed-dotted line}.}
\label{fig:Lin_Lff}
\end{figure*}

Equation (\ref{eq:Lin}) implies that, as a non-turbulent core becomes
unstable and begins to collapse, it follows the trajectory described
by the solid line in the $\LL$-{\it vs.}-$\Sigma$ diagram shown in the
{\it left} panel of Fig.\ \ref{fig:Lin_Lff}. This can be compared to
the locus of a core of the same mass and initial radius, but assuming
it has the free-fall velocity at all times, given by the dotted line
in this figure. This implies that cores that start at rest and have
not yet attained the full free-fall speed will in general appear
sub-virial by an amount that depends on the time elapsed since
collapse started. Finally, note that, in reality, the actual infall
speed must be even lower because, as already pointed out by
\citet{Larson69}, at the early stages of collapse of an object of mass
only slightly larger than the Jeans mass, the thermal pressure is not
negligible, making the collapse slower than free fall.

\subsubsection{Core with turbulence}
\label{sec:turb_case}

Let us now consider that the full 3D velocity dispersion in the core,
$\sigtot$, contains an infall component, given by eq.\ (\ref{eq:vinf})
and a turbulent component $\vturb$, {as discussed in
  Sec.\ \ref{sec:core_evol} which may possibly depend on scale as
\begin{equation}
\vturb = v_0 \left(\frac{R}{R_0}\right)^\eta.
\label{eq:turb_scaling}
\end{equation}
Adding the infall and turbulent components of the velocity dispersion
in quadrature, for such a core the corresponding Larson ratio is
\beq
\Ltot = \frac{\sigtot} {R^{1/2}} = \left\{\frac{6 \pi} {5} G\Sigma \left[1 -
      \left(\frac{\Sigma_0}{\Sigma}\right)^{1/2} \right] +
\frac{\vturb^2} {R}\right\}^{1/2},
\label{eq:Ltot}
\eeq

The {\it left} panel of Fig. \ref{fig:Lin_Lff} also shows, with dashed
lines, the evolutionary path of four cores following eq
(\ref{eq:Ltot}): two with initial turbulent dispersions $\sigma_0 =
\sigma_{\rm in}/2$ (dashed lines) and two with $\sigma_0 = 3
\sigma_{\rm in}$ (dotted lines). In each pair, the upper {curve
  corresponds to a scaling of $\vturb$ with} $\eta = 1/2$, which
corresponds to supersonic turbulent saling \citep[e.g., ][]{Burgers48,
  Passot+88} while the lower one {corresponds to} $\eta = 2$. This
later value does not corresponds to a particular fluid regime, but
allows us to represente the possibility that the turbulent component
dissipates as the core is contracting.

The {\it right} panel of Fig.\ \ref{fig:Lin_Lff} shows the evolution
of all these cases in the Larson diagram, \dvr, with the same line
coding. It can be seen that, in general, the cores move transversely
to the Larson linewidth-size relation, represented by the dotted
curve, terminating their evolution in the upper-left part of the
diagram, as first discussed by \citet{BP11a}, and as do the cores in
our numerical simulations (cf.\ Sec.\ \ref{sec:simulations} below).

A few points are worth noting about these curves. First, it can be
seen that the {contribution} of the turbulent component to the
velocity dispersion is most important at low column densities, for
which the self-gravity-driven (i.e., infall) component is minimum. At
higher column densities, the infall component becomes increasingly
dominant. This is consistent with the result that the Larson ratio
generally exhibits excesses over the equipartition value, or,
equivalently, the virial parameter,
\begin{equation}
 \alpha \equiv \frac{5 \sigmatdmath^2 R}{GM},
\label{eq:alpha_vir}
\end{equation}
is larger than unity, for low column density objects, both in
observational \citep[e.g.,] [] {Barnes+11, Leroy+15} and numerical
\citep{Camacho+16} clump surveys.

Second, we stress that the turbulent component by no means needs to be
interpreted as a supporting mechanism. As found by \citet{Camacho+16},
this component corresponds, in roughly half of the low column density
clouds or clumps, to large scale, {\it external} compressions that are
{\it assembling} these objects, rather than supporting them.

Third, the transition from a domination of turbulent to the
gravitational motions is due to the fact that the two follow different
scalings: while in principle the turbulent velocity dispersion depends
only on scale (eq.\ \ref{eq:turb_scaling}), regardless of column
density \citep[e.g.,] [] {Padoan+16}, the self-gravity-driven ones
depend both on scale and column density (or mass), according to
relation (\ref{eq:larson_general}) \citep{Heyer+09, BP11a, Camacho+16,
  Ibanez+16}.

Fourth, note that clumps or cores that start with a low turbulent
content may, during the initial stages of the contraction, exhibit a
deficiency of the Larson ratio in relation to the equipartition
(virial or free-fall) value. This may explain observations that find
sub-virial cores \citep[e.g.,] [] {Kauffmann+13, Ohashi+16,
  Sanhueza+17}.

Finally, it must be remembered that in the idealized study presented
here, we have considered the monolithic collapse of a single clump
over two orders of magnitude in column density, from values typical of
large GMCs ($\sim 10$ \Msun\ pc$^{-2}$) to those of dense cores ($\sim
10^3$ \Msun\ pc$^{-2}$). In reality, such a range in column density is
not accomplished by the collapse of a single object, but rather,
{through} several stages of fragmentation. Thus, our result should
only be taken as {a first-order approximation to} the effect of
transitioning from an external, turbulence-dominated regime to an
infall-dominated one.

\section{Data}\label{sec:data}

Since we want to understand the relative importance of turbulence
and/or gravity in the process of star formation, we use both
observational and numerical data of massive dense cores to study the
kinematics of the gas before it forms stars.  To accomplish this, we
need to avoid the effects of stellar feedback in both datasets. In the
case of the observational data, even though our cores are located in
massive star-forming regions, we choose without morphological evidence
of perturbations (e.g., jets or free-free emission due to
ultra-compact HII regions). Also, we use ammonia, which might be
destroyed in case of UV stellar feedback \citep[e.g., ][]{Palau+14},
which is expected to trace preferencially dense regions, rather than
molecular flows, and requires large column densities to be
detected. On the numerical side, stars are represented by sink
particles that are allowed to acrette gas from their surrounding, but
no feedback is prescribed in order to preserve the purely
gravitational nature of the velocity field.

\subsection{Observational data}\label{sec:obs}

We use recently published data of cores traced with \nh\ and/or
CH$_3$CN, in order to map cores with high surface densities ($N\ge
10^{21}$cm\alamenos2).  Our sample of clumps and cores is taken from
the works of \citet{Sepulveda+11, Sanchez-Monge+13}, and
\cite{Hernandez-Hernandez+14}.  The first two works determine the
properties from \nh(1,1) and (2,2).  The third sample includes cores
studied in CH$_3$CN.  The clumps and cores of the three samples
include typical star-forming regions distributed all over the Galaxy,
and are not focused on one single cloud, thus avoiding possible biases
due to uncertainties in the distance determination, or to
peculiarities of a given molecular cloud.

Finally, special care was taken in determining the linewidths, sizes
and surface densities using the same method for all the samples. In
particular, \nh(1,1) linewidths were measured using the same routine
in the GILDAS program CLASS, which takes into account the hyperfine
structure of the \nh(1,1) transition. The \nh\ abundance was adopted
to be $4\times10^{-8}$, as an average value of previous works
\citep{Pillai+06, Foster+09, Friesen+09, Rygl+10, Chira+13}.

\subsection{Numerical Simulations}\label{sec:simulations}

In order to numerically investigate the evolution of cores in the
\dvr\ (``Larson'') and \dvrSigma\ diagrams, we performed 5 isothermal
numerical simulations with Gadget-2, a Smoothed Particle Hydrodynamics
(SPH) public code \citep{Springel05} to represent the interior of a
small ($\sim$ parsec size) molecular cloud. We also include sink
particles in the code, as in \citet{Jappsen+05}.
  
Details of the simulations can be found in \citet{BP+15}. Here we just
mention that these simulations were performed using 6 million
particles, with a total mass of 1000~\Msun\ in a cubic open box of
side 1 pc in all simulations but one.  Three of the simulations had an
initially homogeneous density field. We imposed initial velocity
fluctuations using a pure{ly} rotational velocity power spectrum with
random phases and amplitudes, as {in} \citet[][]{Stone+98}, and with a
peak at wavenumber $k_{\rm for} = 4\pi/L_0$,where $L_0$ is the linear
size of the box. No forcing at later times is imposed. The initial
Mach number{s} of these runs were 16, 8 and 4, respectively, so we
label them Run{s} M16, M8 and M4, respectively.

Additionally, we performed two other runs, for which the initial
velocity field is zero. The first one, labeled Run M0-$\rho$K, has
initial density fluctuations with a Kolmogorov-like power spectrum
proportional to $k^{-5/3}$.

Finally, in the run labeled M0-$\rho$P, the initial density field is a
Plummer sphere with $R_c=1$~pc, a size of 5 pc and a total mass of
4500\Msun, again, with zero velocity field.  Since the boxes are all
{strongly} super-Jeans, all of them collapse within {roughly} one
free-fall time.  Furthermore, runs with non-zero velocity field
exhibit a slight initial expansion, since we do not include any
confining external medium.

The global evolution of all these runs can be generically described as
follows: if the initial velocity field is not set to zero, the
velocity fluctuations produce density enhancements while the
turbulence is dissipated. The external parts of the cloud expand
because there is no confining medium, but since the begi{n}ning, the
{bulk of the clouds' mass} start{s to contract}. This contraction,
however, is noticeabl{e} only after some time, depending on the
initial Mach number of the simulation: for run M16, it takes about 1/2
free-fall time $\tff$ for the collapse to begin, $\sim 1/4{\tff}$ for
Mach 8 and {only} a small fraction {of $\tff$} for the Mach 4 run.
Once the initial imposed turbulence is dissipated, all runs proceed to
collapse, lowering their sizes and increasing the velocity dispersion
{(although in this case, this nonthermal velocity dispersion is driven
  by self-gravity and corresponds to chaotic infall rather than actual
  turbulence that can provide support)}. Runs with no initial velocity
field (runs M0-$\rho$K and M0-$\rho$P), only {undergo} this second
part {of the evolution}, with the initial density fluctuations driving
the local centers of collapse {\citep[see, e.g.,] [] {KB00}.}

In order to understand the evolution of these clouds in the
observational diagrams (Larson ratio {{\it vs.}} column density and
velocity dispersion {{\it vs.}} size), we {have} defined 3 regions
(``cores'') {in the simulations} for which we computed their mass,
size and velocity dispersion at every time. These are spheres located
at the center of the computational box, {towards which} the cloud is
collapsing in a chaotic way {due to the turbulent} initial velocity
fluctuations, and {their sizes are defined so that} they contain, at
every time, 10, 25 and 50\%\ of the mass of the whole cloud

The mass and the velocity dispersions are calculated
straightforwardly, the former by just adding the mass of all the SPH
particles {sphere} and the latter as the standard deviation of the
particles' velocities.  }

Finally, the clouds' sizes are calculated as the Lagrangian size,
i.e., as the cubic root of the {sum of the} volume{s of}
all the SPH particles:

\begin{equation}
    R = V_{\rm tot}^{1/3} = \left[m_{\rm sph} \sum_i{1\over \rho_i}
    \right]^{1/3}
\label{eq:R}
\end{equation}
where $\rho_i$ is the density of the $i^{\rm th}$ particle.

\section{Results}\label{sec:results}

\subsection{Non-existence of a \dvr\ relation}\label{sec:results:_obs}

As mentioned in \S\ref{sec:intro}, \citet{BP11a} showed that current
{observational} datasets using high-mass MC core observations in
different {kinds of} regions do not exhibit the \citet{Larson81}
\dvr\ relation.  Instead, they occupy the upper-left corner of the
plane.  Although such cores can be expected to have large column
densities because they belong to massive star-forming regions, most of
the data in the compilations discussed by \citet{BP11a} did not have
estimations of the actual masses of the cores. Thus, the proposal that
cores with larger column densities have larger velocity dispersions
and smaller sizes still requires further testing using observational
data that provide column density determinations {that are} independent
{of} the velocity dispersion measurements.

\begin{figure*}
	\includegraphics[width=\columnwidth, angle=0]{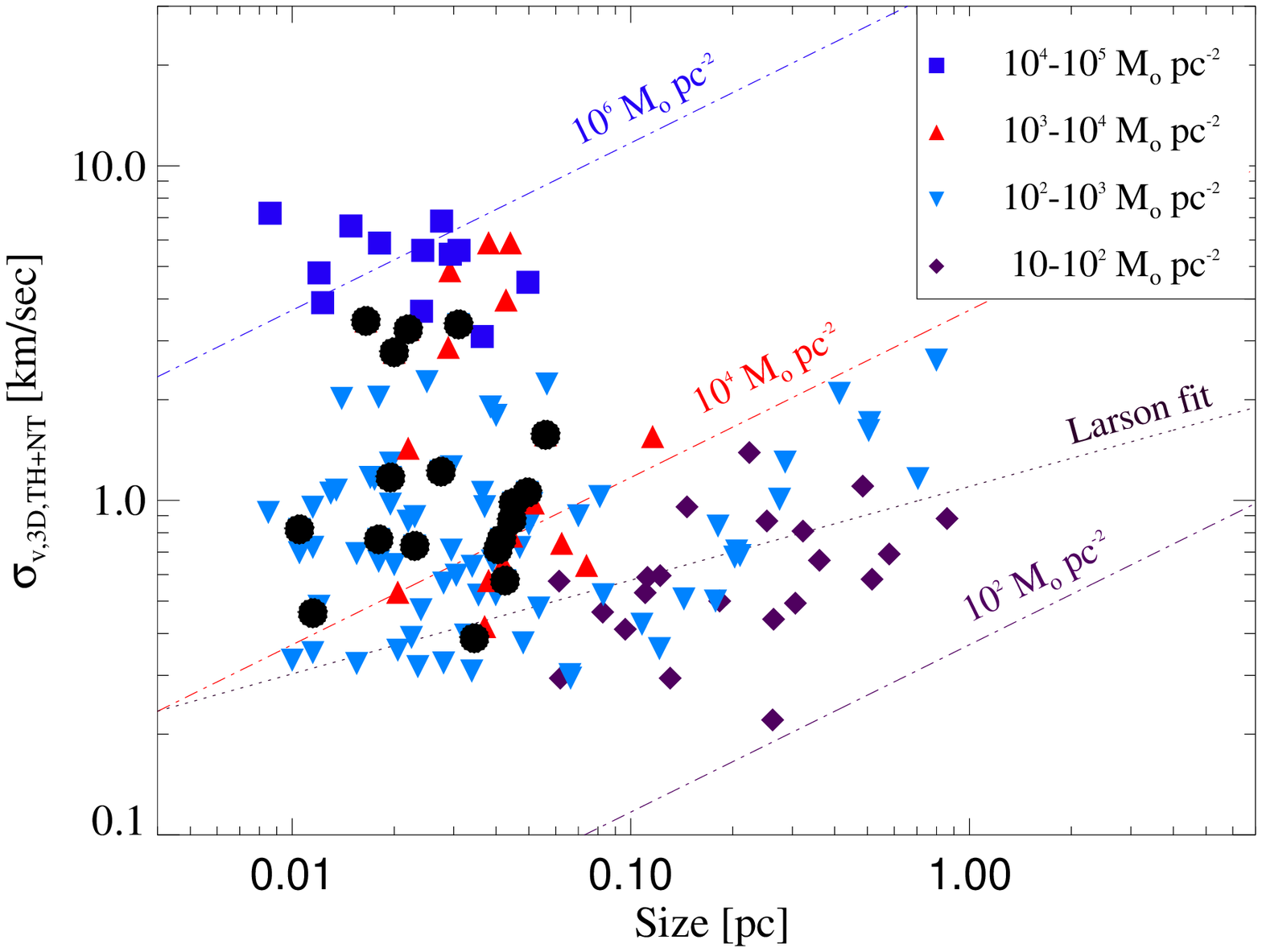}
	\includegraphics[width=\columnwidth, angle=0]{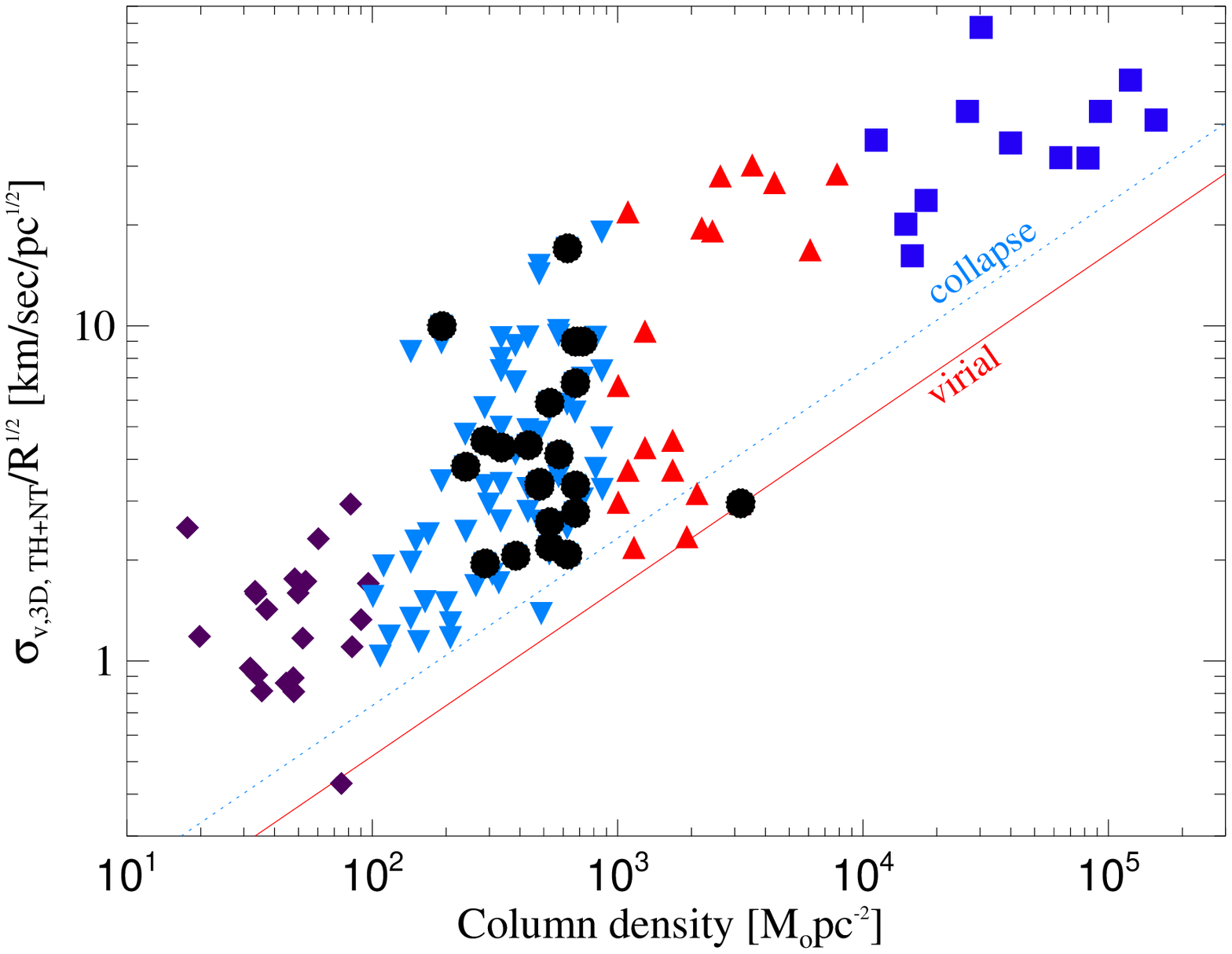}
\caption{Right: Velocity dispersion {\it vs.} size for the compiled
  observational sample.  Colors and symbols denote different ranges in
  column density, according to the labels. Lines of constant column
  density if non-thermal motions are driven by self-gravity are shown
  as dot-dashed lines.  The original fit provided by \citet{Larson81}
  is shown with the solid line. In addition, we overlap with a dark
  solid circle those cores that have been catalogued as quiescent
  {and} starless. Left: same sample, in the Larson ratio $\LL$ {\it
    vs.} column density $\Sigma$ space.}
    \label{fig:larson_obs}
\end{figure*}

In Figure~\ref{fig:larson_obs}a we {plot} our observational data
sample in the Larson plane \dvr. W<e use \sigmatd, defined as
$\sqrt{3} ~ \sigma_{\rm v, obs}$, assuming that the three-dimensional
velocity dispersion \sigmatd\ is intrinsically isotropic, while the
observed velocity dispersion $\sigma_{\rm v, obs}$ is only the
projection of the former along the line of sight.  The dataset is
colored according to the column density of each core: purple diamonds
represent the lowest column density points, ranging between 10 and
\diezala{2}\Msun pc\alamenos2; light blue triangles represent the
column density range \diezala{2}-\diezala3 \Msun pc\alamenos2; red
triangles, \diezala{3}-\diezala4\Msun pc\alamenos2, and blue squares,
\diezala{4}-\diezala6\Msun pc\alamenos2. In addition, we {highlight
  some} cores with a dark circle, {indicating those} cores {that are}
catalogued as quiescent {and} starless by \citet{Sanchez-Monge+13}.

{In this plot,} the black {dotted} line represents the fit {obtaned}
by \citet{Larson81}, with a slope of 0.38, while the colored dashed
lines represent lines of constant column density, assuming
equipartition ($\Ek = |\Eg|$) and that the cloud {is spherical and}
has uniform density, so that
\begin{equation}
  \Eg = -{3\over 5} {G M^2\over R}.
  \label{eq:gravity_cst}
\end{equation}
Although there is substantial scatter due to the observational
uncertainties, the cores follow a clear trend: higher-column density
clumps have larger velocity dispersions and smaller sizes.

In Figure \ref{fig:larson_obs}b we plot the ratio $\LL \equiv$
\larsonratio\ against the column density $\Sigma$ for the same sample
show{n} in Fig.~\ref{fig:larson_obs}a.  Note that, if the clouds
actually followed the two Larson relations ($\sigmatdmath\propto
R^{1/2}$ and $n\propto R^{-1}$) simultanesouly, all data points
{would} collapse to a single point {in this plot} (within the
intrinsic scatter), as pointed out by \citet{Heyer+09}. The {red}
solid line {in Fig.\ \ref{fig:larson_obs}b} represents the velocity
dispersion of clouds in virial equilibrium, while the dashed line
represents clouds in monolithic free-fall (cf.\ \S
\ref{sec:core_evol}). We notice that, with some scatter, cores tend to
follow a relation $\sigmatdmath/R^{1/2} \propto \Sigma^{1/2}$,
although for our sample, the cores have sistematically
larger-than-virial values of the Larson ratio. Note that the clouds
analyzed by \citet{Heyer+09} also exhibit larger-than-virial values,
although those authors argued that this may have been due to their
clouds' masses possibly having been underestimated by factors $\sim
2$.

The fact that {our} data points are {in general} located above the
virial-equilibrium {and free-fall} line{s in the \dvrSigma\ plane}
would {normally be interpreted as implying} that the objects are not
bound. However, this argument is somehow contradictory, since both our
sample and the data of \citet{Heyer+09} contain high-column density
objects, and thus they are likely to be strongly gravitationally
bound.  Moreover, note that, as evident in Fig.~\ref{fig:larson_obs}b,
the free-fall line lies slightly above the virial equilibrium one;
that is, the kinetic energy associated with free-fall is slightly
larger than that associated with virial equilibrium.  We will explore
this point in \S\ref{sec:discussion}.

\subsection{Evolutionary trends of collapsing
cores} \label{sec:results_sims}

In order to interpret the observational data, in particular the
apparent average kinetic energy excess over equipartition of the whole
sample, which is not predicted by the analytical model of
Sec.\ \ref{sec:core_evol}, we now turn to the evolution of the cores
in the numerical simulations.

\begin{figure*}
  \includegraphics[width=\columnwidth, angle=0]{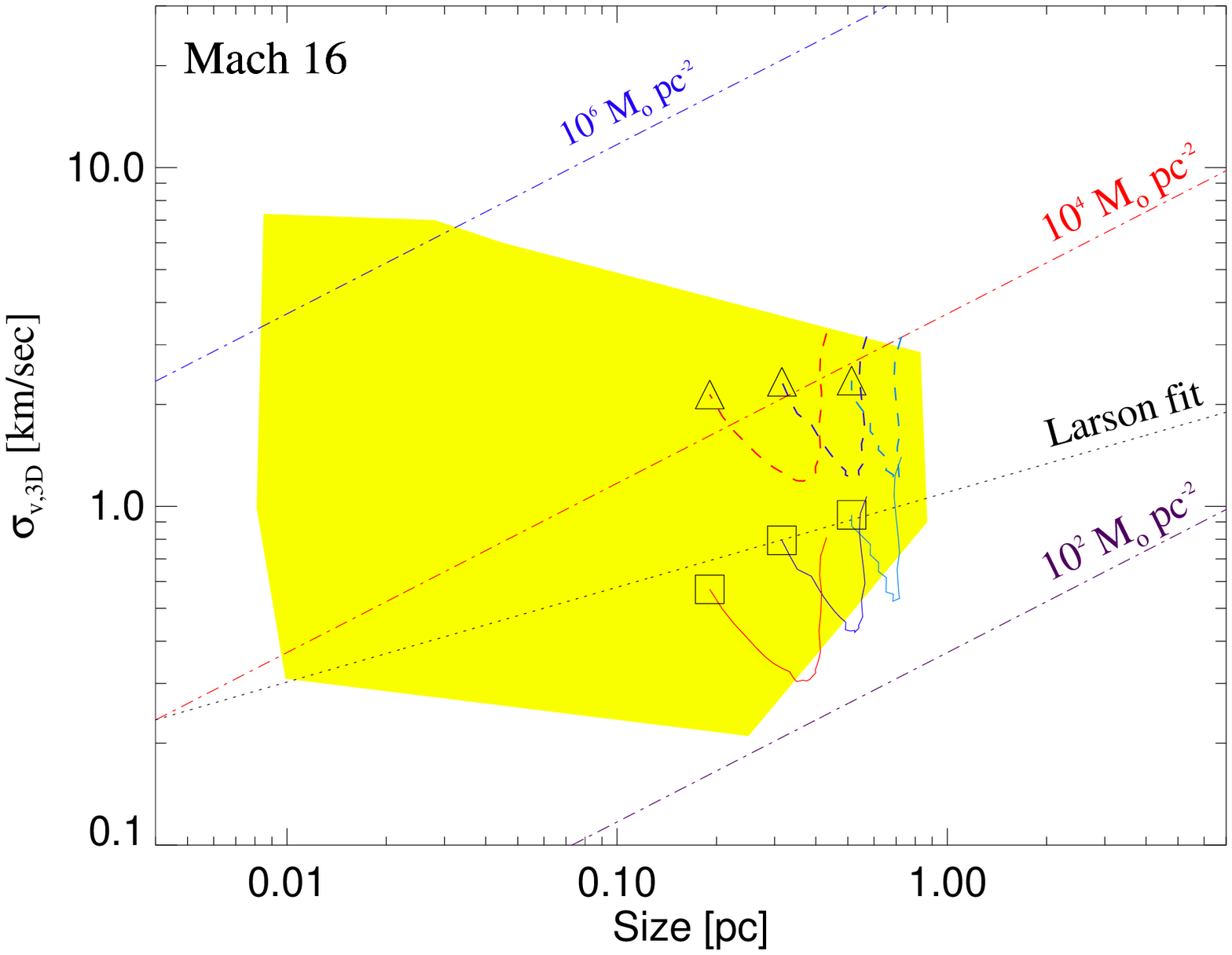}
  \includegraphics[width=\columnwidth, angle=0]{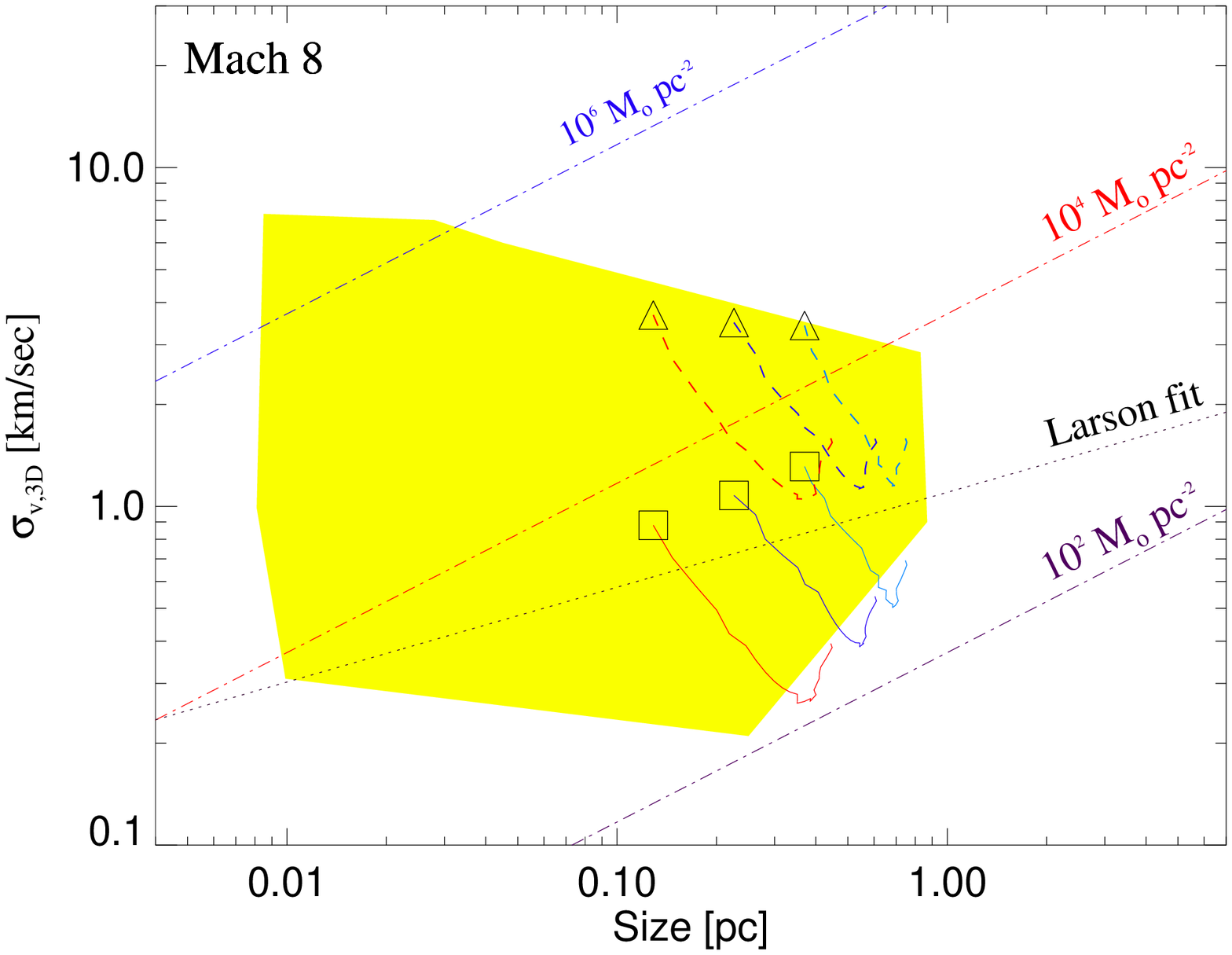}
  \includegraphics[width=\columnwidth, angle=0]{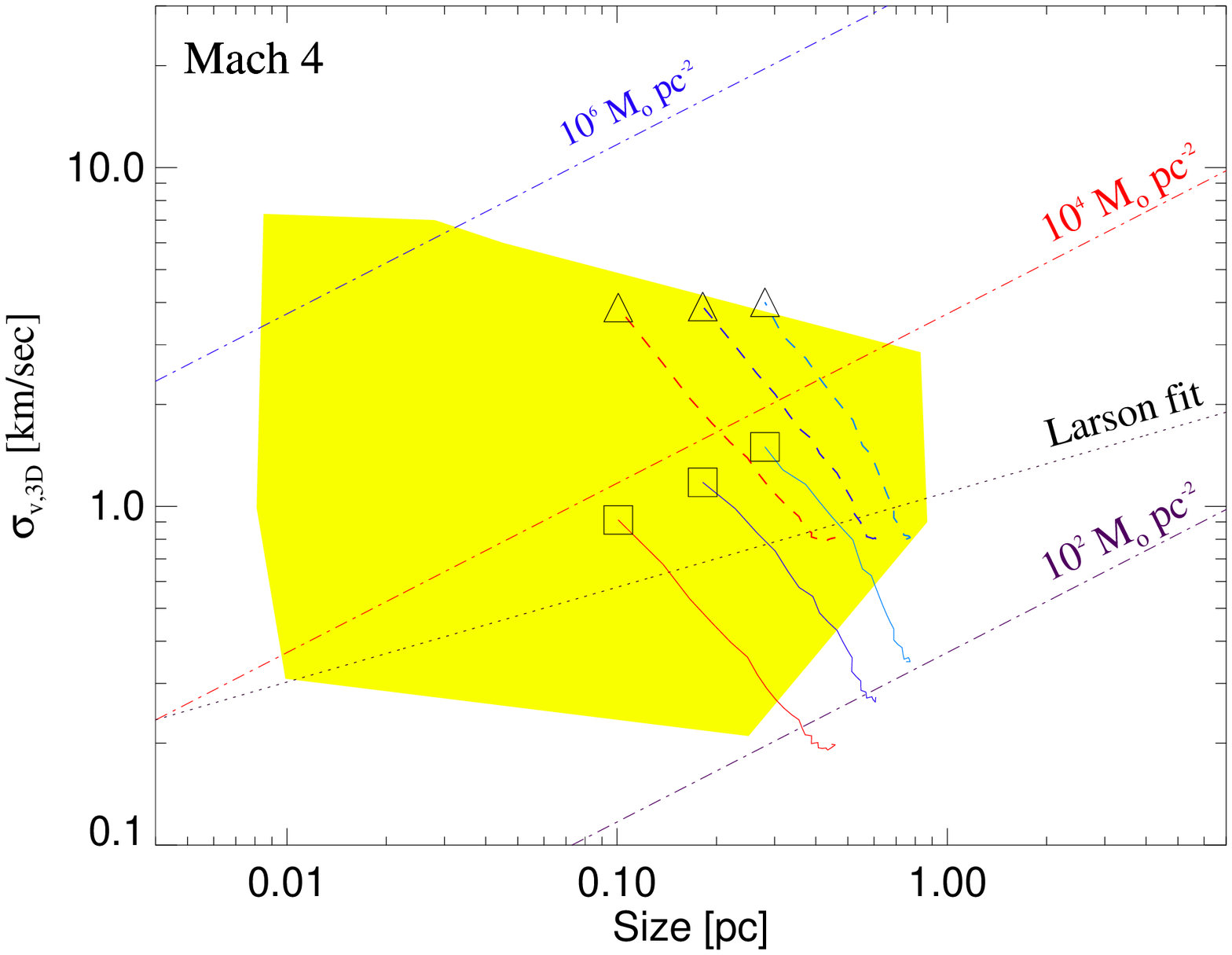}
  \includegraphics[width=\columnwidth, angle=0]{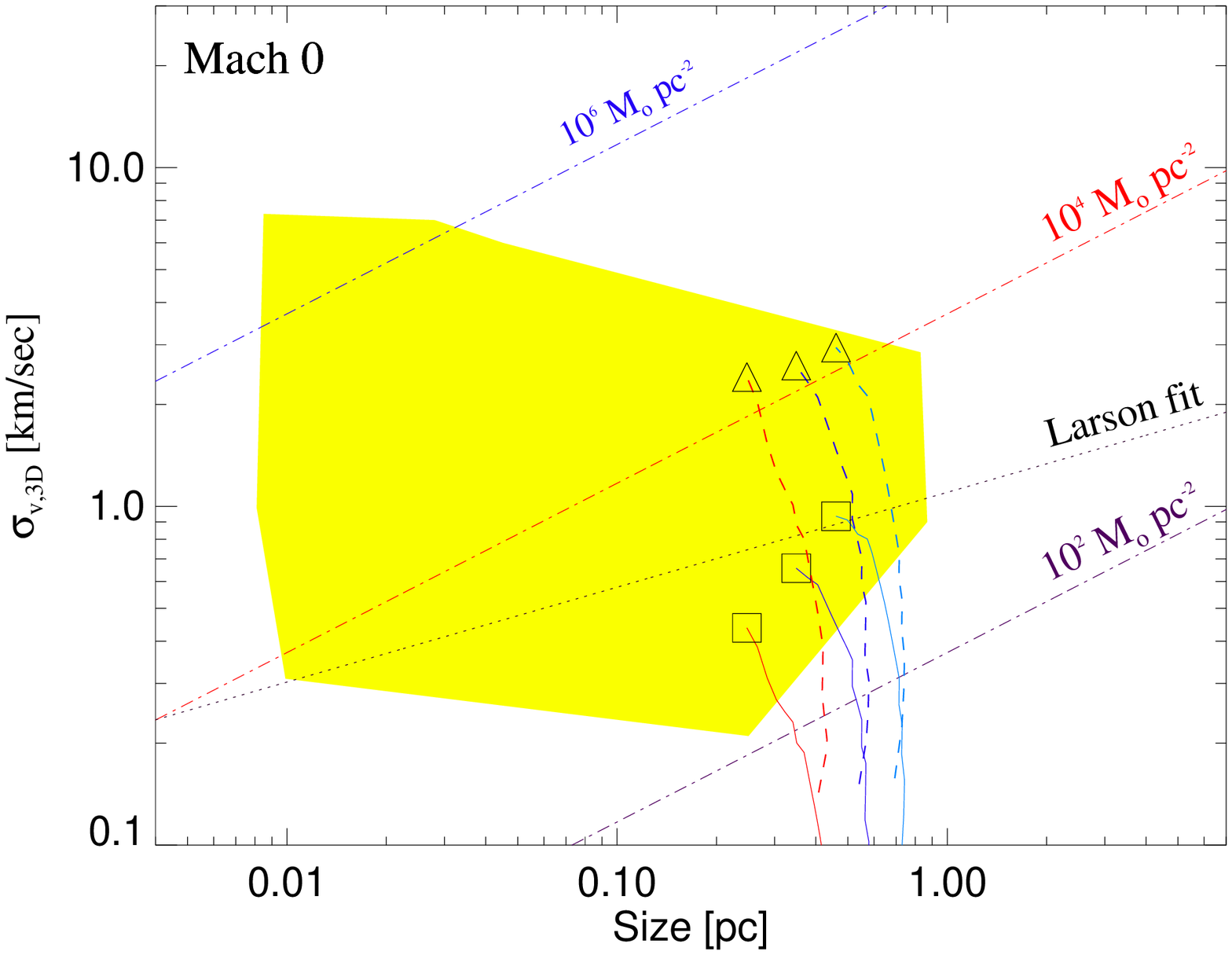}
   \includegraphics[width=\columnwidth, angle=0]{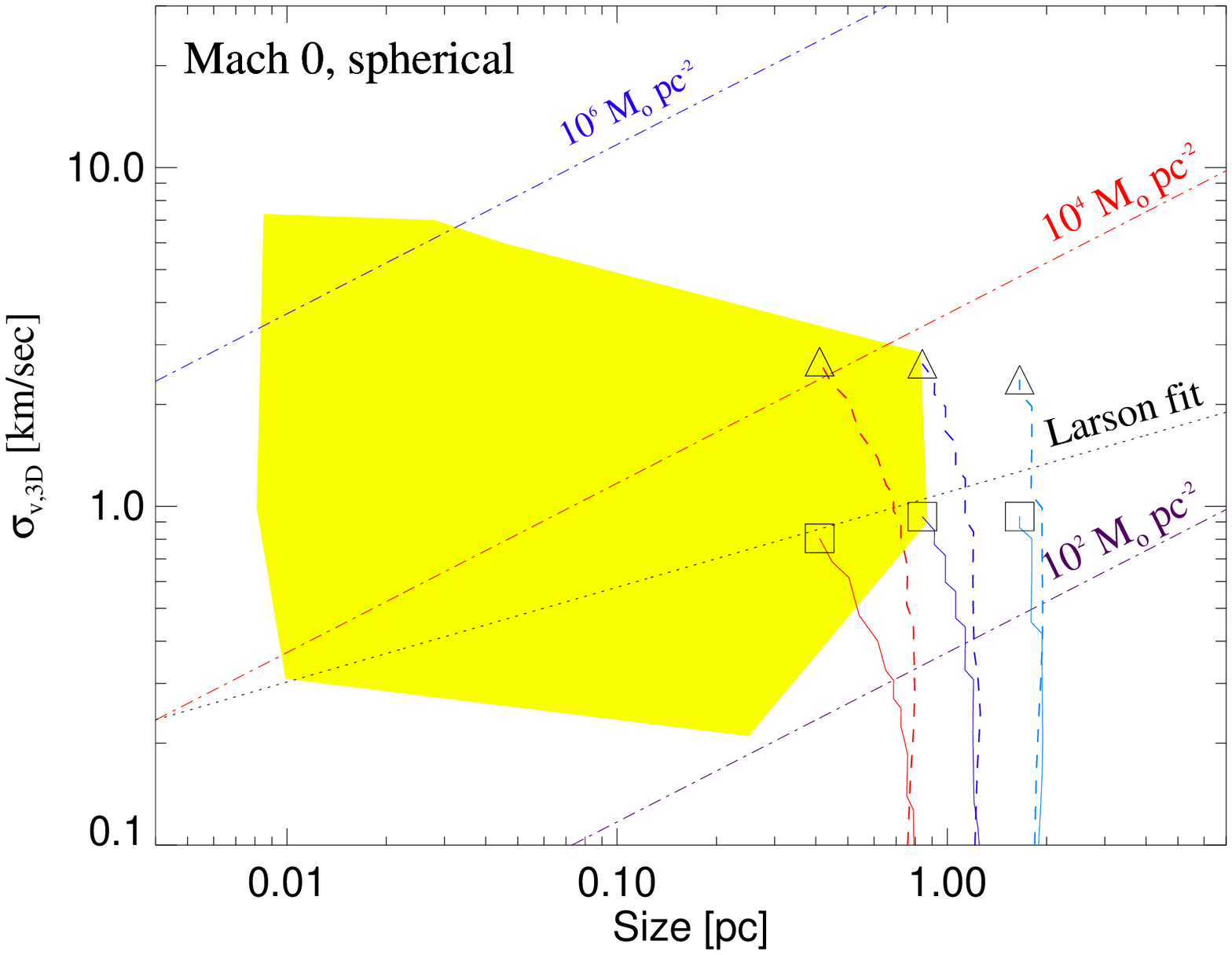}
   \caption{Dashed lines: evolution of the {cores in the} simulations
     in the Larson plane.  Panels are organized as follows: (a) Run
     with initial Mach number 16. (b) Mach 8, (c) Mach 4, (d) Mach 0
     with Kolmogorov density fluctuations, and (e) Mach 0, Plummer
     profile. Each curve represents aregion that contains 50\%
     (rightmost curve), 25\% and 10\% (leftmost curve) of the mass at
     each time.  Dashed lines are lines where energy equipartition is
     achieved at different column densities (see eq.~(\ref{eq:Lff}).
     Solid lines: evolution of the cores in the simulations, but
     considering the correction factor defined in
     \S\ref{sec:correction}.  Triangles and squares denote the last
     time in the evolution }
    \label{fig:larson_sim}
\end{figure*}

In Fig.~\ref{fig:larson_sim}, we show the evolution of {the} cores in
the Larson plane.  As in Fig.~\ref{fig:larson_obs}b, the dotted line
represents Larson's original relation, with a slope of 0.38, while the
dot-dashed lines represent lines of constant column densit{y}.  For
reference, the yellow area denotes the {region occupied by the} cores
shown in Fig.~\ref{fig:larson_obs}a.  In each panel, the evolution of
the numerical cores is indicated with {\it dashed} lines, with red,
dark purple and cyan representing the cores containing 10, 20 and
50\%\ of the total mass in the box, respectively. The solid lines are
discussed in the next section.  The length of these trajectories
corresponds to one free-fall time at the initial mean density for runs
M16, M8 and M4, and to 0.8 free-fall times for the zero-velocity runs
(M0-$\rho$K and M0-$\rho$P), since these simulations contain very
dense regions where the timestep becomes too small, slowing the
simulation down towards the end of the evolution. The triangles at the
end of these lines indicate the place in which each core terminates
its evolution.  In all simulations, the cores with the higher masses
have larger radii, so that the rightmost curves represent the
evolution of the core with 50\%\ of the mass, while the leftmost ones
correspond to those with only 10\% of the mass.

The evolution of the cores in the Larson diagram can be described as
follows: Runs M16 and M8 initially exhibit a decrease in the velocity
dispersion.  The former {does so} at nearly constant size, while the
latter {does so while} slightly lowering its size. {This decrease}
corresponds to the initial period during which the excess turbulent
energy dissipates.  But, as gravity takes over, the sizes start
decreasing {more rapidly} and the velocity dispersion starts
increasing again, causing the cores to move towards the upper left in
the Larson diagram. In the case of runs with small (M4) or no
(M0-$\rho$K and M0-$\rho$P) initial turbulence, the cores only undergo
the second stage, lowering their size while increasing their velocity
dispersion.

\begin{figure*}
  \includegraphics[width=\columnwidth, angle=0]{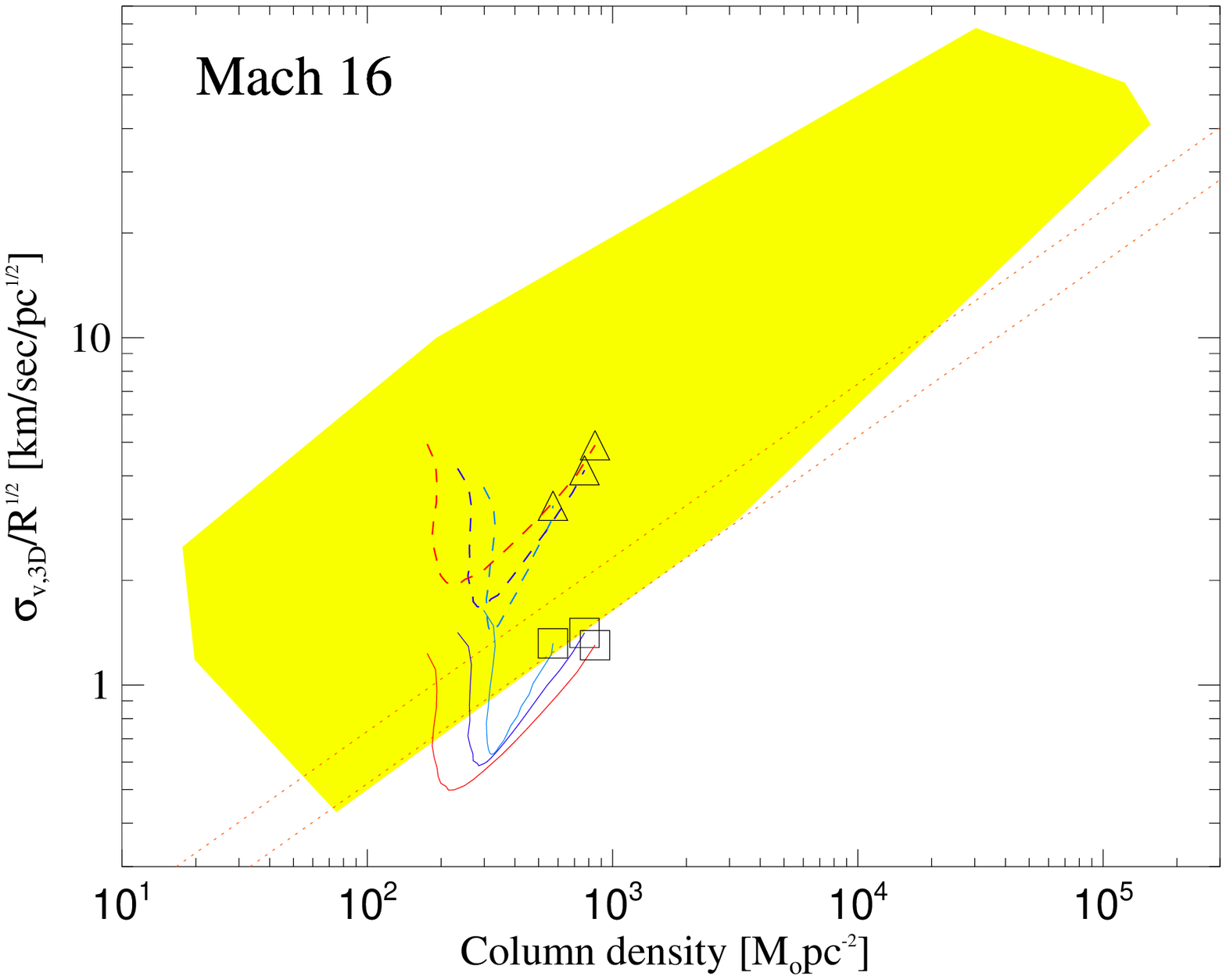}
  \includegraphics[width=\columnwidth, angle=0]{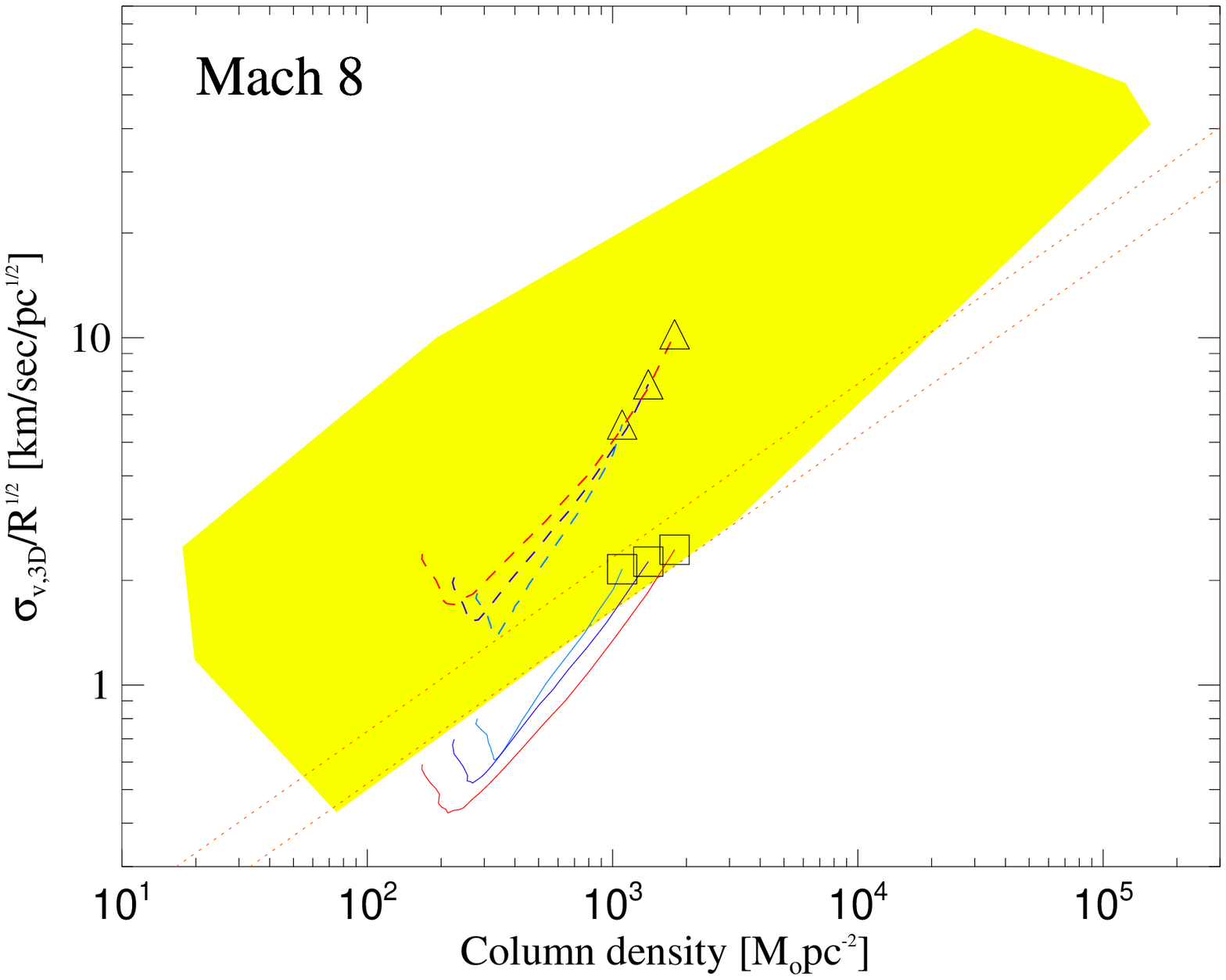}
  \includegraphics[width=\columnwidth, angle=0]{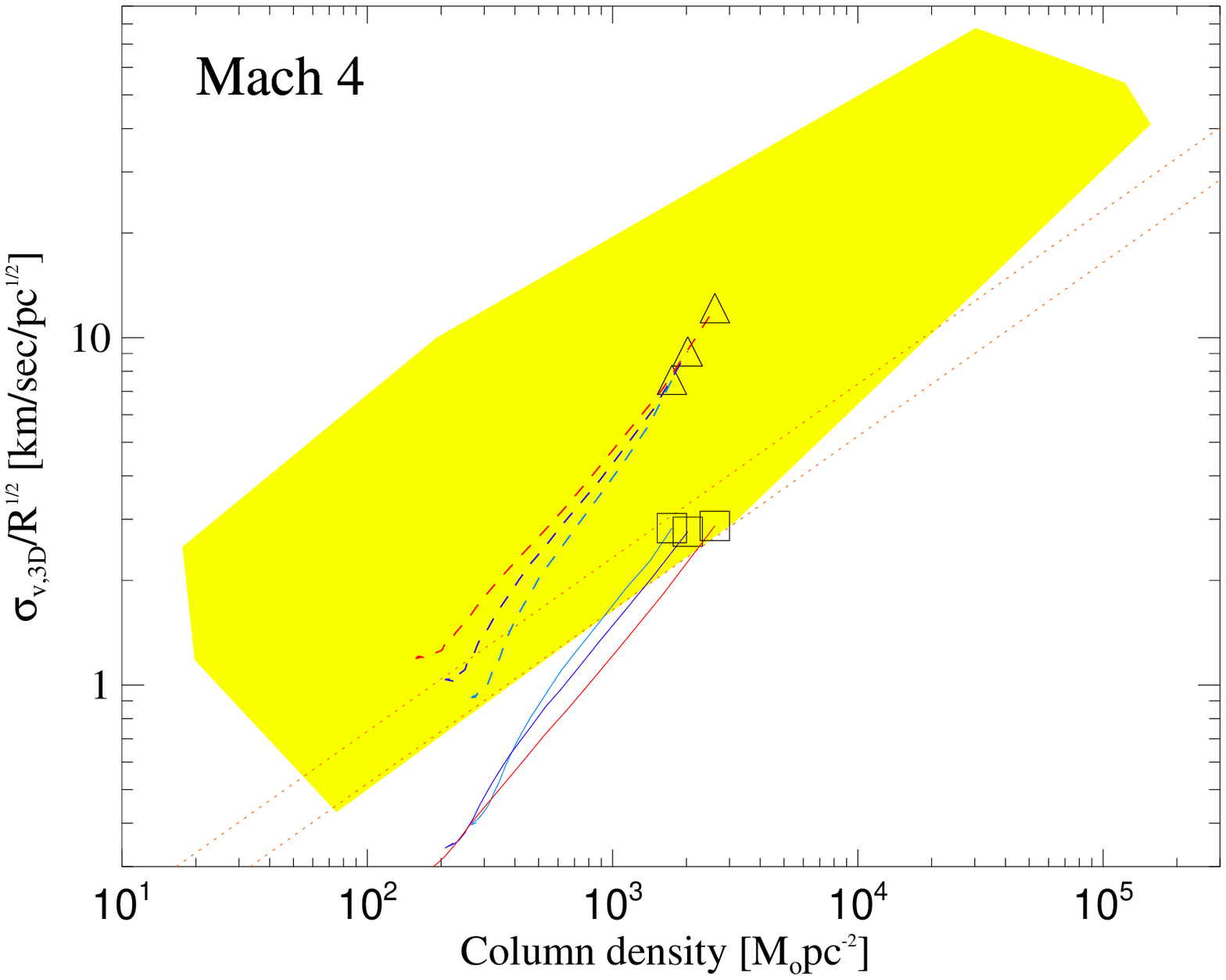}
  \includegraphics[width=\columnwidth, angle=0]{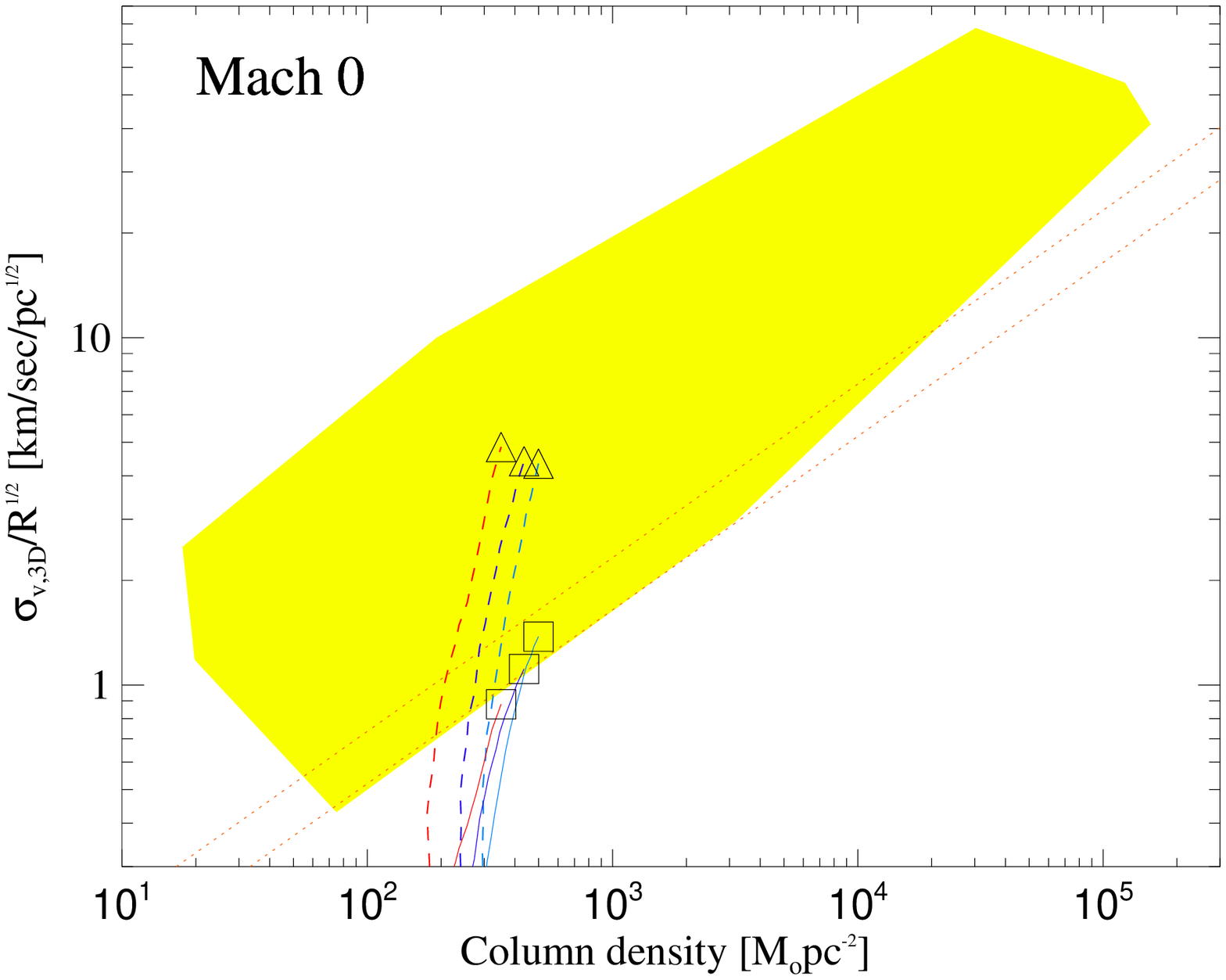}
   \includegraphics[width=\columnwidth, angle=0]{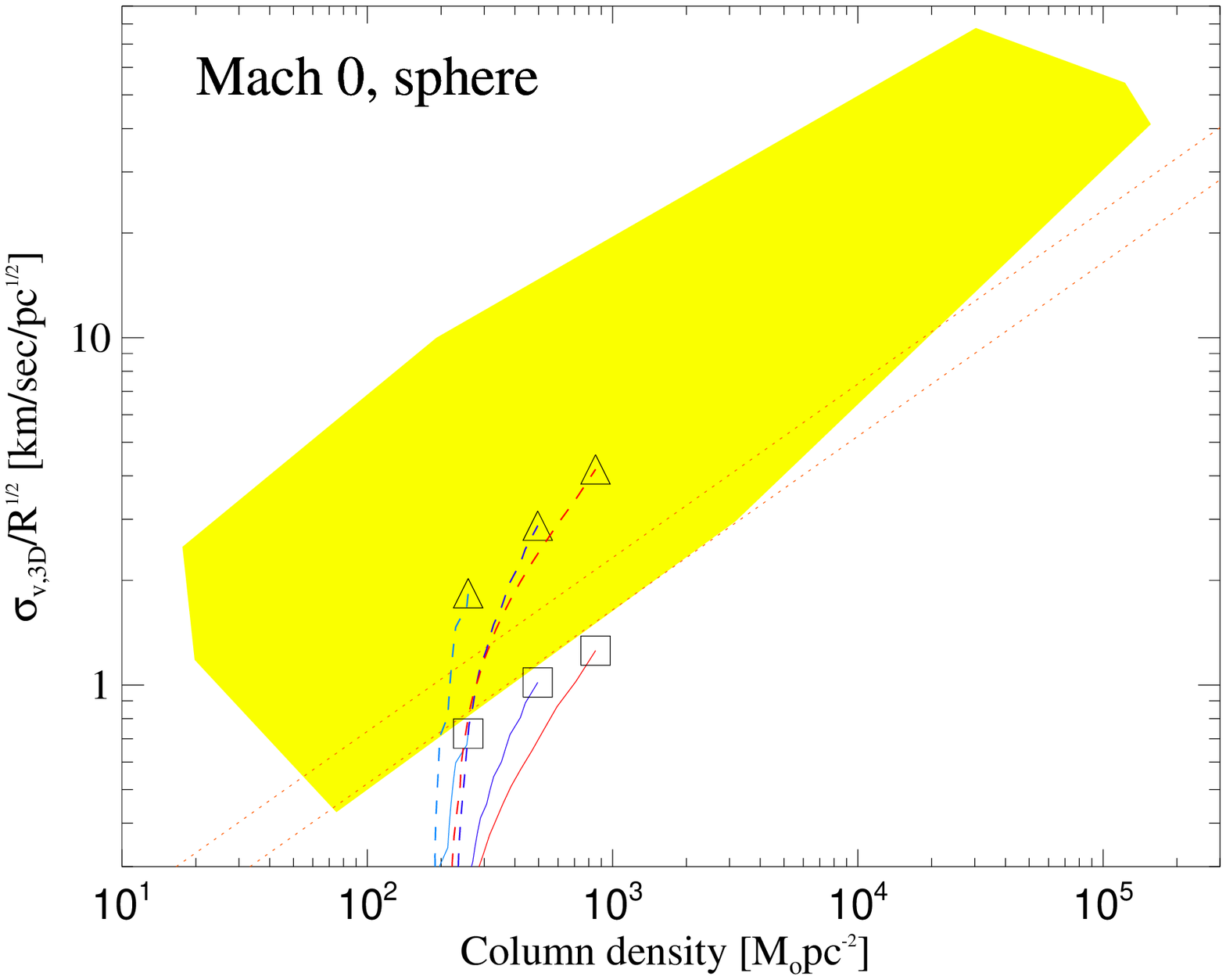}
\caption{Dashed lines: evolution of the cores in the simulations in
  the Larson ratio-{\it vs.}-column density plane.  Panels and
  notation, as in Fig.~\ref{fig:larson_sim}.  The two lines in this
  figure denote the locus of the Virial and energy equipartition of a
  sphere with constant density.}
    \label{fig:heyer_sim}
\end{figure*}

In Fig.\ \ref{fig:heyer_sim} we show the Larson ratio {$\LL = $}
\larsonratio \ vs. the mass column density $\Sigma$ {of the cores in
  our} simulations. The lower and upper red dotted lines with slope
1/2 represent the locus of sphere with uniform density in virial
equilibrium and in energy equipartition at the indicated column
density, respectively.  The yellow shaded area again denotes the
region occupied by our observational data (see
Fig.~\ref{fig:larson_obs}a).

The evolution of our cores in the Larson ratio-column density $\Sigma$
diagram is represented, again, by the dashed lines: red, dark purple
and cyan representing the cores containing 10, 20 and 50\%\ of the
total mass in the box, respectively. As in the previous figure, the
solid lines are discussed in the next section. As can be expected,
runs with an excess of initial kinetic energy (M16, M8) lower their
$\LL$ ratio during the initial dissipation period, but as soon as
gravity takes over, they increase it again, while their column density
also increases.  Instead, runs with low initial turbulence (M4) or
without it (M0-$\rho$K and M0-$\rho$P) increase their Larson ratio
$\LL$ while they also increase their column density, in agreement with
the prediction of the analytical model.

What seems to be surprising in Fig.~\ref{fig:heyer_sim} is that, even
though the cores we show are all collapsing, they {terminate their
  evolution} exhibit{ing} Larson ratios in excess of the equipartition
and virial {values corresponding to their column density}.  Thus,
although our cores are not confined by any external medium and are
collapsing, if only the information of this plot were provided, they
would be naively interpreted {either} as expanding cores, or as cores
that require an external confining pressure in order to be in virial
equilibrium and to avoid their expansion.

Another point to notice is that the final column densities of the cores
that would be inferred {from their} final position in the Larson diagram
({comparing with the constant-column density lines in} Fig.\
\ref{fig:larson_sim}) are large{r than} their actual final column
densities (as denoted by the abscissa of the triangles in Fig.\
\ref{fig:heyer_sim}). For instance, the Larson diagram would imply a
column density of the order of $\sim$\diezala4 \Msun pc\alamenos2 for
{the cores from} run M16 (upper-left panel in
Fig.~\ref{fig:larson_sim}), {although they} actually end {their
evolution} with a column density of $\Sigma\sim 10^3$\Msun pc\alamenos2
({see the} upper-left panel in Fig.~\ref{fig:heyer_sim}).  In the
following section we will show that {these} results can be easily
explained when the true gravitational energy of the cores, i.e., the
gravitational energies considering the complex internal density structure.

\section{Apparently supervirial collapsing cores: need for a correction
  factor}\label{sec:correction}

It is clear from the previous section that observed cores do not
follow Larson's velocity dispersion-size scaling relation, and may
exhibit apparent excesses of kinetic energy when compared to the
gravitational energy of a spherical cloud with constant density of the
same mass and size, as is implicitly done by the standard virial
equilibrium and equipartition lines drawn in
Figs.~\ref{fig:larson_obs}b and \ref{fig:heyer_sim}.

In the standard picture, clouds/cores with such excesses {are
  interpreted} either as {expanding} \citep[e.g.,][]{Dobbs+11}, or as
confined by an external medium \citep[e.g.,][]{KM86, BertoldiMcKee92,
  Field+11}.  {A frequent} explanation of the origin of this apparent
kinetic energy excess in those cores is that they belong to massive
star forming regions, and thus they may be subject to {kinetic energy
  injection from} stellar feedback.  However, {at least in the case of
  the cores in our observational sample,} the observed molecules
(NH$_3$ and CH$_3$CN) are thought to not be seriously affected by
stellar feedback: ammonia is destroyed by radiation and winds from
stars, while CH$_3$CN requires large column densities to be detected,
and thus it is unlikely that it comes from the outflows.  In addition,
{even} those cores that are {known} to be starless and quiescent in
our sample (denoted by overlaid filled circles in
Fig.\ \ref{fig:larson_obs}b) also exhibit an excess of kinetic energy.
In this case, how should we interpret this apparent excess?

To answer this question we turn to the simulations, in which the cores
are collapsing and nevertheless still exhibit kinetic energies in
excess of the corresponding gravitational energy as given by
eq.~(\ref{eq:gravity_cst}).  The solution to this apparent
contradiction is {actually} rather simple: $\Eg$, the gravitational
energy of a sphere with constant density {as given by
  eq.\ (\ref{eq:gravity_cst}),} is only a lower-limit to the magnitude
of the actual gravitational content of the cloud.  In practice,
centrally concentrated density structures generally have a more
negative gravitational energy than that of uniform-density structures
of the same mass and size \citep[see, e.g., ][]{BertoldiMcKee92}; that
is,
\begin{equation}
  |E_{\rm g,real}| > |\Eg |,
\end{equation}
where  $\Eg$ is the approximation {defined in}  eq.\
(\ref{eq:gravity_cst}), and  
\begin{equation}
  E_{\rm g,real} = - {1\over 2}\int_V \rho\phi dV
  \label{eq:gravityAA}
\end{equation}
is the true gravitational energy,  in which 
\begin{equation}
  \phi = G \int_{\rm all\ space} {\rho(x^\prime)\over |{\bf x} - {\bf x^\prime} |
    } dV
  \label{eq:potencial}
\end{equation}
is the {\it total} gravitational potential, i.e., the potential
produced by the mass distribution over all space, and not only inside
the volume $V$ over which the integral in eq.\ (\ref{eq:gravityAA}) is
performed.  Therefore, by using the actual gravitational potential
instead of eq.\ (\ref{eq:gravity_cst}) for comparison with the kinetic
energy, we can avoid the two approximations: the assumption of a
constant density, and the neglect of the mass external to the cores.
However, because both the Larson diagram and the $\LL$--$\Sigma$
diagram involve the velocity dispersion, it is more convenient to
introduce a correction factor to the velocity dispersion, which can be
used in both diagrams.

Since at their late stages of evolution our cores are collapsing,
either because the
turbulent energy has been dissipated, or was not included at all,
the non-thermal motions have a purely gravitational origin, and thus
equipartition between kinetic and gravitational energy should
 hold at
sufficiently late times during the evolution (cf.\ Sec.\
\ref{sec:core_evol}).  Thus we write
\begin{equation}
  \Ek =  -E_{\rm g,real}.
  \label{eq:equiparticion}
\end{equation}
Defining the factor 
%
 $$\Gamma \equiv {\Eg\over W_{\rm cl}},$$
%
and subtituting it in eq. (\ref{eq:equiparticion}), we can compare the
modified kinetic energy of each core with the
gravitational {energy} of a sphere with constant density {of the same
mass and ``size'' $R$ (cf.\ eq.\ [\ref{eq:R}]) as the actual
cloud}.  We obtain

\begin{equation}
  \Gamma \Ek = -\Eg.
\end{equation}
 From this equation it is clear that we need to multiply the measured
 ordinate axes of Figs.\ \ref{fig:larson_sim} and
 \ref{fig:heyer_sim} by the factor $\Gamma$ in order to appropriately
 compare the kinetic energy of the cores to the gravitational energy
 of a homogeneous sphere, and so we define:
\begin{eqnarray}
 \left(\sigmatdmath\right)_{\rm corr} &\equiv&
 \Gamma\ \sigmatdmath \label{eq:corrected_ylarson}
 \\
\LL_{\rm corr} &\equiv& \Gamma \LL,
 \label{eq:corrected_yheyer}
\end{eqnarray}
with $\LL$ as defined by eq. (\ref{eq:LL}).  In
Figs.\ \ref{fig:larson_sim} and \ref{fig:heyer_sim} we {thus}
respectively show, with solid lines, the velocity dispersion {{\it
    vs.}} size and the Larson ratio {\it vs.} column density diagrams
with the corrected $y$ axes, as given by
eqs.~(\ref{eq:corrected_ylarson}) and (\ref{eq:corrected_yheyer}).  In
both cases, as expected, the values of the {ordinates} are {now}
smaller than in the uncorrected case (dashed lines).  But more
importantly, now the values are consistent: the collapsing cores do
not show an excess of kineitc energy compared to the gravitational
content, and instead they terminate their evolution near the
equipartition/virial lines in Fig.~\ref{fig:heyer_sim}.  Also, the
final column densities predicted by the equipartition lines at
constant column density in Fig.~\ref{fig:larson_sim} (dot-dashed
lines) are now in better agreement with the final column densities
shown in Fig.~\ref{fig:heyer_sim}.  These results reflect the fact
that collapse does induce {virial-like}
equipartition \citep{VS+07, BP11a}.

Finally, we also note that the {location} of our cores in the diagrams
depends on their evolutionary stage, as discussed in \S
\ref{sec:core_evol}.  Although we {have} not run the simulations more
than one free-fall time because the timestep of the simulation becomes
too short as collapse proceeds, the evolutionary tracks, as well as
the simple model {discussed} in \S\ref{sec:core_evol}, suggest that
further evolution in the Larson's ratio-column density diagram is
expected to proceed roughly along the virial/equipartition lines.  If
a similar correction is applicable to the observational sample, then
its locus in the Larson ratio-column density diagram would be {in
  closer agreement} with equipartition/virial balance, suggesting that
{the} internal motions {of the cores in the sample} are indeed
dominated by gravity.  Furthermore, the evolution {in} the Larson
diagram should be expected to be oblique to the lines {of} constant
column density, with a slope that should approach $-1/2$, since

\begin{equation}
  \sigma \propto \sqrt{\Sigma R} \propto
         {
  R^{-1/2}}
\end{equation}
{at} constant mass, as  {is the case of}
the cores  {discussed} in this section {and
in Sec.\ \ref{sec:core_evol}}. 

\section{Discussion} \label{sec:discussion}

Traditionally{, the apperent near-virialization} exhibited by clouds
{and their substructures (clumps and dense cores) has been interpreted
  as} a manifestation that {all the structures in this hierarchy} are
supported against their self-gravity by strongly supersonic
turbulence, {and that, when this} turbulence dissipates at the
smallest scales, the cores {can then proceed} to collapse \citep[see,
  e.g., the reviews by] [] {Larson81, VS+00, MK04, BP07,
  McKeeOstriker07, BT07, HF12, Dobbs+14, Donkov+11, Donkov+12,
  Veltchev+16}.  However, it is difficult to understand how the {many}
energy-injection mechanisms could adjust themselves to provide just
the right amount of energy to the turbulence to keep the structures
nearly virialized {{\it at all scales}}. Some authors have proposed,
on the basis of idealized models for the star formation rate
\citep[e.g.,] [] {Krumholz+06, Goldbaum+11}, that the feedback from
stellar sources internal to the clouds can self-regulate to attain
near virialization. However, evolutionary models \citep{Zamora+12,
  ZV14}, as well as numerical simulations \citep[e.g.,] [] {VS+10,
  Dale+12, Dale+13a, Dale+13b, Colin+13} do not show a trend towards
local virialization, but rather, towards local destruction of the
star-forming sites.

{On the other hand, it is also sometimes suggested that feedback
from sources external to the clouds can drive the turbulence at all
scales within the clouds} \citep[e.g.,] [] {Padoan+16}. {The
underlying assumption here is}
that this is a natural consequence of the turbulent
cascade, as the energy spectrum of Burgers-like strongly supersonic
turbulence should approach the form $E(k) \propto k^{-2}$, where $k$ is
the wavenumber. This spectrum implies a velocity dispersion-size scaling
relation of the form $\sigma \propto R^{1/2}$, similar to Larson's
scaling, and thus strongly supersonic turbulence has been proposed as
the origin of the Larson scaling. However, as shown in several
observational and numerical studies \citep[] [this work; although see
Padoan et al.\ 2016] {Heyer+09, BP11a, Leroy+15, Camacho+16}, the Larson
scaling is {\it not} universal, and instead the Larson ratio $\LL$,
which should be constant for strongly supersonic
turbulence, is actually dependent on the column density as
$\Sigma^{1/2}${, suggesting that the process does not rely only on
the turbulent cascade}.

More specifically, within the scatter, cores tend to be organized by
column density along lines with slope 1/2 in the Larson
$\sigmatdmath$-$R$ diagram, with high-column density cores being
located on the upper-left part of the diagram, and low-column density
cores appearing at lower velocity dispersions at a given value of $R$.
Moreover, regardless of the observational uncertanties, not all cores
with the same column density will necessarily be located exactly at
the same position in the {\dvrSigma} diagram, since their exact
position on the diagrams depends on their evolutionary stage, as the
results from \S\S \ref{sec:core_evol} and \ref{sec:results_sims} show.

The observational sample considered here also shows {a correlation
  between} the Larson ratio $\LL$ {and} the column density {in general
  agreement with} eq.~(\ref{eq:larson_general}), although {on average}
it exhibits somewhat supervirial values when the gravitational energy
{of the sample cores} is computed using eq.\ (\ref{eq:gravity_cst})
\citep[e.g.,] [] {Heyer+09, Leroy+15}, which assumes that the cloud is
a uniform{-density} sphere.  Although we cannot discard stellar
feedback in the observations, the facts that {a)} our sample was
observed with tracers that should not be strongly affected by
feedback; {b)} the subsample of quiescent starless cores exhibits the
same behavior, and {c)} simulations of collapsing cores also exhibit
overvirialization, lead us to conclude that such {apparent}
overvirialization {may} be spurious.  Indeed, we find that, when the
kinetic energy is corrected by a factor equal to the ratio of the
actual gravitational energy to that given by
eq.\ (\ref{eq:gravity_cst}), then the apparent excess of kinetic
energy essentially disappears.

On the other hand, surveys of massive star-forming cores sometimes
exhibit {sub}virial values of the Larson ratio, a result which has
often been interpreted as the cores being supported by some form of
energy other than the turbulent pressure, such as the magnetic
pressure \citep[e.g.,] [] {Kauffmann+13, Ohashi+16, Sanhueza+17} , or
else as the cores being already in full-blown collapse, i.e., at the
full free-fall speed, rather than the lower infall speed given by
eq. (\ref{eq:vinf}).  However, because the infall speeds are also
nonthermal motions, full-blown collapse should correspond to a virial
ratio (cf.\ eq.\ [\ref{eq:alpha_vir}]) $\alpha \sim 2$, not to $\alpha
< 1$.  Although {magnetic support for these cores} is indeed a
plausible explanation, we have shown that, when a core begins its own
local collapse, it may start with subvirial velocities {if its intial
  turbulent energy is sufficiently low}, approaching {the} virial
value from below.  This constitutes an alternative plausible
explanation for the observations of subvirial cores.

The notion that the nonthermal motions observed in MCs are produced by
the collapse itself is contrary to the widespread notion that
turbulence is the main physical process controlling the internal
dynamics of MCs, providing support to MCs and causing their
fragmentation.  Although the turbulent density fluctuations { in the
  diffuse medium} may very well play a crucial role in the formation
of the seeds of what eventually will grow as cores via instabilities,
the fluctuations produced self-consistently by this mechanism are
generally not strong enough to become locally Jeans unstable, until
global collapse has significantly reduced the mean Jeans mass in the
cloud \citep[e.g., ][]{KI02, Heitsch+05, CB05, VS+07, HH08}. Moreover,
observational data show that fragmentation levels within dense cores
do not correlate with the intensity of the observed non-thermal
motions \citep{Palau+15}, again suggesting that they actually do not
consist of random turbulent fluctuations. Nevertheless, the
fluctuations produced by the growth of seeds triggered by the global
collapse do contain a significant fraction of chaotic motions due to
the turbulent background, and so theese fluctuations are far from
being ordered and monolithic. Indeed, \citet{Heitsch+09} showed that
clouds in a state of hierarchical gravitational collapse do exhibit
line profiles similar to observed MCs {in} the structure of the CO
lines, the supersonic widths, and their core-to-core velocity
dispersion.  However, the net average component of the velocity field
continues to be dominated by global contraction, as shown by studies
of the dense regions in simulations of driven, isothermal turbulence,
which indicate that the overdensities tend to have a negative net
velocity divergence (i.e., a convergence) {\citep[e.g.] [] {VS+08,
    Gonzalez-Semaniego+14, Camacho+16}.}  Thus, the density
fluctuations are in general contracting, rather than being completely
random with zero or positive net divergence, as would be necessary for
the bulk motions to exert a ``turbulent pressure'' capable of opposing
the self-gravity of the overdensities. If the non-thermal motions in
the clumps and cores do not exert a turbulent pressure capable of
providing support against the self-gravity of the structures, then MC
models based on the competition between gravity and turbulent support
{may need to} be revisited \citep[e.g.,] [] {MT03, KM05, HC08, HC11,
  Hopkins12}.

\section{Conclusions}\label{sec:conclusions}

In this paper we have used observational data and numerical
simulations to show that the scaling relation between velocity
dispersion and size does not hold when column densities {spanning a
  large dynamic range} are {considered}.  {In this case}, cores with
large column densities tend to be located in the upper-left corner of
the {Larson} velocity dispersion-size diagram. Using numerical
simulations, we show{ed} that{,} as cores collapse, their sizes become
smaller and their column densities larger, and thus, their evolution
tends to follow lines oblique to the \sigmatd$\propto R^{1/2}$
relation.

Additionally, we showed analytically that, as cores evolve from when
they first detach from the global flow and begin their local collapse,
they may exhibit subvirial velocities if their internal turbulent
component is initially low enough. This is because, when the core
first starts to collapse locally, its gravitationally-driven speed
starts out from zero, and takes a finite amount of time to reach the
full free-fall speed. This behavior was observed also in the numerical
simulations that start with little or no turbulent energy, and it may
explain recent observations of apparently sub-virial cores
\citep[e.g., ] [] {Kauffmann+13, Ohashi+16, Sanhueza+17}.

We also showed that, although the observed cores were selected to
avoid stellar feedback, they appear to be supervirial.  However, this
feature is likely due to the fact that, rather than comparing the
kinetic energy to the actual gravitational content of the core, $W$,
in practice it is customary to use the gravitational energy of a
sphere with constant density as a proxy for $W$. This proxy actually
provides only a lower limit (in absolute value) to the actual value of
$W$. We show that, indeed, this is the case in our simulations:
collapsing cores do exhibit over-virial velocities at the end of their
evolution, and thus end up in the virial/equipartition zone.  Thus,
the virial parameter computed using the uniform sphere approximation
should be taken with caution when interpreting their observational
data.

This work then provides support to the {scenario} that non-thermal
motions in MCs have largely a gravitational origin and are dominated
by infall motions. {Because of} the rapid dissipation of turbulence,
{the conversion of the infall motions into random, truly turbulent
  motions that can} oppose the collapse that produces them in the
first place {does not appear feasible}.  In a future contribution, we
plan to {further explore this possibility.}

\section{Acknowledgments}

This work was supported by UNAM-PAPIIT grant number IN110816 to JBP
and CONACYT grant 255295 to E.V.-S.  In addition, J.B.P. acknowledges
the hospitality of the Institute for Theoretical Astrophysics of the
University of Heidelberg, as well as UNAM's DGAPA-PASPA Sabbatical
program. He also is indebted to the Alexander von Humboldt Stiftung
for its valuable support.  RSK acknowledges support from the Deutsche
Forschungsgemeinschaft in the Collaborative Research Center SFB 881
“The Milky Way System” (subprojects B1, B2, and B8) and in the
Priority Program SPP 1573 “Physics of the Interstellar Medium” (grant
numbers {\tt KL 1358/18.1}, {\tt KL 1358/19.2}). RSK furthermore
thanks the European Research Council for funding in the ERC Advanced
Grant STARLIGHT (project number {\tt 339177}).  Numerical simulations
were performed in the supercomputer Miztli, at DGTIC-UNAM.  We have
made extensive use of the NASA-ADS database.

{}


\end{document}